\documentclass{emulateapj}

\usepackage{graphicx}

\newcommand{\beq}{\begin{equation}}
\newcommand{\eeq}{\end{equation}}

\newcommand{\teff}{$T_{\rm eff}$}

\def\earth{\hbox{$\oplus$}}
\def\arcmin{\hbox{$^\prime$}}

\def\solar{\mbox{$_{\normalsize\odot}$}}

\newcommand{\lsim}{\ \raise
-2.truept\hbox{\rlap{\hbox{$\sim$}}\raise5.truept\hbox{$<$}\ }}
\newcommand{\gsim}{\ \raise
-2.truept\hbox{\rlap{\hbox{$\sim$}}\raise5.truept\hbox{$>$}\ }}
\newcommand{\simsim}{\ \raise
-2.truept\hbox{\rlap{\hbox{$\sim$}}\raise5.truept\hbox{$\sim$}\ }}

\newcommand{\Msun}{M$_\odot$}

\slugcomment{Accepted for publication in The Astrophysical Journal}
\shorttitle{Age and age-spread of pre--main-sequence stars in the young LMC cluster LH~95}
\shortauthors{Da Rio, Gouliermis, Gennaro}

\begin{document}

\title{A new method for the assessment of age and age-spread of pre--main-sequence stars in Young Stellar Associations
of the Magellanic Clouds\altaffilmark{$\star$,} \altaffilmark{$\dagger$}}
\author{Nicola Da Rio\altaffilmark{*}, Dimitrios A. Gouliermis, Mario Gennaro\altaffilmark{*}}
\affil{Max-Planck-Institut f\"{u}r Astronomie,  K\"{o}nigstuhl 17, D-69117 Heidelberg, Germany}
\email{dario@mpia-hd.mpg.de}
\email{dgoulier@mpia-hd.mpg.de}
\email{gennaro@mpia-hd.mpg.de}

\altaffiltext{$\star$}{Based on observations made with the NASA/ESA {\em
Hubble Space Telescope}, obtained at the Space Telescope Science
Institute, which is operated by the Association of Universities for
Research in Astronomy, Inc. under NASA contract NAS 5-26555.}

\altaffiltext{$\dagger$}{Research supported by the German Aerospace
Center (Deutsche Zentrum f\"{u}r Luft und Raumfahrt).}

\altaffiltext{*}{Member of IMPRS for Astronomy \& Cosmic Physics at the University of Heidelberg}


\begin{abstract}
We present a new method for the evaluation of the age and age-spread among pre--main-sequence (PMS)
stars in star-forming regions in the Magellanic Clouds, accounting simultaneously for photometric errors, unresolved binarity,
differential extinction, stellar variability, accretion and crowding. The application of the method is performed with the statistical construction
of synthetic color-magnitude diagrams using isochrones from two families of PMS evolutionary models. We  convert each isochrone into 2D probability distributions of artificial PMS stars in the CMD by applying the
aforementioned biases that dislocate these stars from their original CMD positions. A maximum-likelihood technique is
then applied to derive the probability for each observed star to have a certain age, as well as the best age for the
entire cluster.
We apply our method to the photometric catalog
of $\sim$~2~000 PMS stars in the young association LH~95 in the Large Magellanic Cloud, based on the deepest HST/ACS
imaging ever performed toward this galaxy, with a detection limit of $V\sim28$, corresponding to $M\sim0.2$~M{\solar}.
We assume the Initial Mass Function and reddening distribution for the system, as they are previously derived by us.
Our treatment shows that the age determination is very sensitive to the considered grid of evolutionary models and
the assumed binary fraction. The age of LH~95 is found to vary from 2.8~Myr to 4.4~Myr, depending on these
factors.
We evaluate the accuracy of our age estimation and
we find that the method is fairly accurate in the PMS regime, while the precision of the measurement of the age is
lower at higher luminosities. Our analysis allows us to disentangle a {\sl real} age-spread from the apparent CMD-broadening
caused by the physical and observational biases.
We find that LH~95 hosts an age-spread well represented by a gaussian distribution with a FWHM of the order of 2.8~Myr to 4.2~Myr depending on the model and binary fraction.
We detect a dependence of the average age of the system with stellar mass. This dependence does not appear to have
any physical meaning, being rather due to imperfections of the PMS evolutionary models, which tend to predict lower
ages for the intermediate masses, and higher ages for low-mass stars.
\end{abstract}

\keywords{
stars: formation ---
stars: pre-main sequence ---
Hertzsprung-Russell and C-M diagrams ---
open clusters and associations: individual (LH~95) ---
methods: statistical}


\section{Introduction}

The Large and Small Magellanic Clouds (LMC, SMC), the dwarf metal-poor companion-galaxies to our own, are well-suited
environments for the study of {\sl resolved} star formation at low metallicity. Indeed, both galaxies are characterized by an
exceptional sample of star-forming regions, where hydrogen has been already started to be ionized by the UV-winds of
massive stars, the {\sc Hii} regions \citep[][]{henize56, davies76}. A plethora of young stellar associations and clusters are
embedded in these regions  \citep[e.g.,][]{bica95, bica99}, giving evidence  of recent (or current) clustered star formation.
For many years, studies of the bright stellar content of these systems has provided and still provide unprecedented
information on massive stellar evolution at low metallicities \citep[see, e.g.,][]{walborn95, walborn10}. On the other hand,
their faint stellar content, which is still in the pre--main-sequence (PMS) phase, became only recently accessible
{\sl exclusively} through imaging with the {\sl Hubble Space Telescope} \citep[see e.g.,][]{gouliermis06,nota06}.

The identification  of the low-mass PMS stars of star-forming regions in the Milky Way is quite demanding, requiring various
techniques such as variability surveys, spectroscopy, detection of emission lines and excess due to accretion
\citep[e.g.,][]{briceno07}, in order to decontaminate the observed stellar samples from the evolved populations of the
Galactic disk. This is not a problem for the observational study of low mass population of  young clusters in the Magellanic Clouds, where memberships can be assigned simply from the position of the stars in the color-magnitude diagram (CMD) alone. This is guaranteed by the negligible distance spread of the LMC or SMC field, in comparison to the distance of the two galaxies.

Naturally, this observational advantage has important implications in our general understanding of clustered star formation and in consequence of PMS
evolution, the Initial Mass Function (IMF) and dynamical evolution of young star-forming clusters. Clearly, in order to detect low-mass stars in the MCs, imaging with high resolution and high sensitivity in optical wavelengths is required, and the {\sl Advanced Camera for Surveys} (ACS) onboard the HST has been so far an ideal instrument for this purpose. Indeed, deep imaging at a detection limit of $m_{\rm 555}
\simeq 28$ with ACS/WFC of the young LMC cluster LH~95 \citep{lucke70} proved for the first time that young MCs
stellar systems host exceptional large numbers of faint PMS stars \citep{gouliermis07}, providing the necessary  stellar numbers to
address the issue of the low-mass IMF in star-forming regions of another galaxy \citep{darioLH95}.

PMS stars remain in the contraction phase for several times 10~Myr, and therefore they are chronometers of star
formation and its duration. However, the estimate of stellar ages in PMS regime from photometry is not without difficulties. This is because the observed fluxes are affected by several physical and observational effects. While these stars are optically visible, they are still surrounded by gas and dust from their formation, typically in circumstellar disks. Irregular accretion, due to instabilities in these disks, produces variability in the brightness of these stars.
Moreover, PMS sources exhibit periodic fluctuations in light produced by rotating starspots.
The variable nature of PMS stars dislocates them from their theoretical positions in the CMD. This effect in addition to observable
biases, such as crowding and unresolved binarity, results in a broadening of the sequence of PMS stars in the CMD. As a
consequence the measurement of the age of the host system and any intrinsic age-spread, which may indicate continuous
star formation, cannot be derived from simple comparisons of PMS isochrones with the observed CMD.
This problem is particulary critical for our understanding of the star formation processes in every region. In particular, \citet{naylor09} showed that neglecting unresolved binaries may lead to an underestimation of the age of a young cluster of a factor 1.5-2. Concerning the star formation duration, the apparent luminosity spread in the CMDs, if not disentangled from a possible intrinsic age spread, leads to an overestimation of the star formation timescale.

With spectroscopic and variability surveys with HST in star-forming regions of the MCs being practically impossible, the only
way to disentangle the age and the duration of star formation in such regions from the rich samples of PMS stars is with the
application of a statistical analysis, from a comparison between the observed CMDs and simulated ones.

In this study we present a new maximum-likelihood method, for the derivation of the ages of PMS stars and the identification of any age-spread among them.
We apply our method to the rich sample of PMS stars detected with ACS in  LH~95  to determine the
age of the stars from their positions in the CMD in a probabilistic manner, accounting for the physical and observational
biases that affect their colors and magnitudes. Our data set is presented in \S~\ref{s:dataset}, and the evolutionary models
to be used for our synthetic photometry are described in  \S~\ref{section:theorisochrones}. In order to achieve the most
accurate results we further refine our previous selection of PMS stars members of the system in \S~\ref{section:fieldsub_refinement}.
In \S~\ref{section:age_estimation_main} we construct modeled 2D density distributions of PMS stars in synthetic CMDs by
applying the considered observational and physical biases to the evolutionary models, and we apply our maximum-likelihood
technique for the derivation of the age of the system. In  \S~\ref{section:deriving_age_spread} we use our method to investigate
the existence of an age-spread in LH~95, and we discuss our findings in \S~\ref{section:discussion}. The summary of this study
is given in \S~\ref{section:summary}.

\section{The data set} \label{s:dataset}

In this study we make use of the photometry of the young association LH~95, obtained from deep {\em Hubble Space Telescope}
(HST) imaging with the Wide-Field Channel of the {\sl Advanced Camera for Surveys} (ACS) in the filters F555W and F814W,
roughly corresponding to standard $V$- and $I$-bands respectively (program GO-10566; PI: D. Gouliermis). The reduction of the
images and their photometry
is discussed in \citet{gouliermis07}, and is presented in its entirety by \citet[][from hereafter Paper~I]{darioLH95}.
The photometric completeness reaches the limit of 50\% at $m_{\rm 555}\simeq27.75$~mag and $m_{\rm 814}\simeq27$~mag,
reduced due to crowding by
$\sim$~0.5~mag in the central area occupied by the association. A control field $\sim$~2\arcmin\ away from the system was
also observed with an identical configuration. It has been used in Paper~I to isolate the true stellar population of the association
from that of the general LMC field by performing a statistical field subtraction on the observed color-magnitude diagram (CMD).
The HST observations of the two fields and in the two bands were not consecutive. In particular, the  $I-$band imaging of LH95 was carried out 4 days after the $V-$ band. This will be crucial in the analysis we present in this work to model the observed effect of rotational variability.

The derived cluster-member population includes 2~104 stars located in the PMS and Upper main-sequence (UMS) region of the CMD
with corresponding masses as low as $\sim 0.2$~\Msun. In Paper~I we isolate the central area of the observed field-of-view
(FoV), which covers a surface of 1.43 arcmin$^2$ ($\sim260$~pc$^2$) corresponding to about 12.6\% of the whole FoV, as the most representative
of the association LH~95. This area, which encompass the vast majority of the observed PMS population, is referred to as the
``{\sl central region}''.

Given low density of stars in LH95 (of the order of 10 stars/pc$^2$ down to $0.3-0.2$~M$_\odot$), it can be considered a loose association. In our previous work three denser substructures - referred to as ``\emph{subcluster A, B, C}'' - were identified. They include about half of the stellar population, while another half is distributed within the central region.
In Paper~I we roughly estimated the dynamical properties of the central region as well as the three subclusters. This was done estimating the total stellar mass, extrapolating our IMF for masses below our detection limit, and assuming a velocity distribution for the stars of the order of $\sim2$~km/s. For the entire region, the crossing time is comparable with the age of the system, and it is few times shorter when computed singularly for each of the three subclusters. It was still unclear whether these stellar agglomerates were formed independently on each other, from separate star forming events, or if their spatial distribution was simply caused by a clumpy distribution of the ISM in the parental cloud, during a single star formation event. In Section \ref{section:spatial_variability} we investigate this, excluding the presence of a spatial variability of the stellar ages.

\section{Description of the evolutionary models} \label{section:theorisochrones}

For the purposes of our study we calculate a set of evolutionary models for PMS stars with the use of the
{\sc FRANEC} (Frascati Raphson Newton Evolutionary Code) stellar evolution code \citep{chieffi89, deglinnocenti08}.
Opacity tables are taken from
\cite{ferguson05} for $\log T[K] < 4.5 $ and \cite{iglesias96} for higher temperatures. The equation of state
(EOS) is described in \cite{rogers96}. Both opacity tables and EOS are calculated for a heavy elements
mixture equal to the solar mixture of \cite{asplund05}. Our models are then completely
self-consistent, with an \emph{unique} chemical composition in the whole structure.

The value of the mixing length parameter, $\alpha$, has been chosen by fitting a solar model, obtained to be $\alpha = 1.9$.
Helium and metal content of the models are those best representative for the LMC, i.e. $Y=0.25$ and $Z=0.01$ respectively.
The evolutionary tracks have been calculated for a range of masses between 0.2 and 6.5 $M_\odot$. We have interpolated a
dense grid of isochrones from the evolutionary tracks, which covers ages from 0.5 to 200~Myr, with steps of 0.1~Myr below 10~Myr, with steps of 0.5~Myr between 10 and 20~Myr, with steps of 1~Myr between 20 and 50~Myr isochrones, and with steps of 5~Myr above 50~Myr.
For ages $\tau<1.6$~Myr the isochrones do not reach the main
sequence (MS) for the highest masses in our sample, while at $\tau\sim50$~Myr the models cover also post-MS evolutionary
phases for intermediate masses.

\begin{figure}[t!]
\includegraphics[width=\columnwidth]{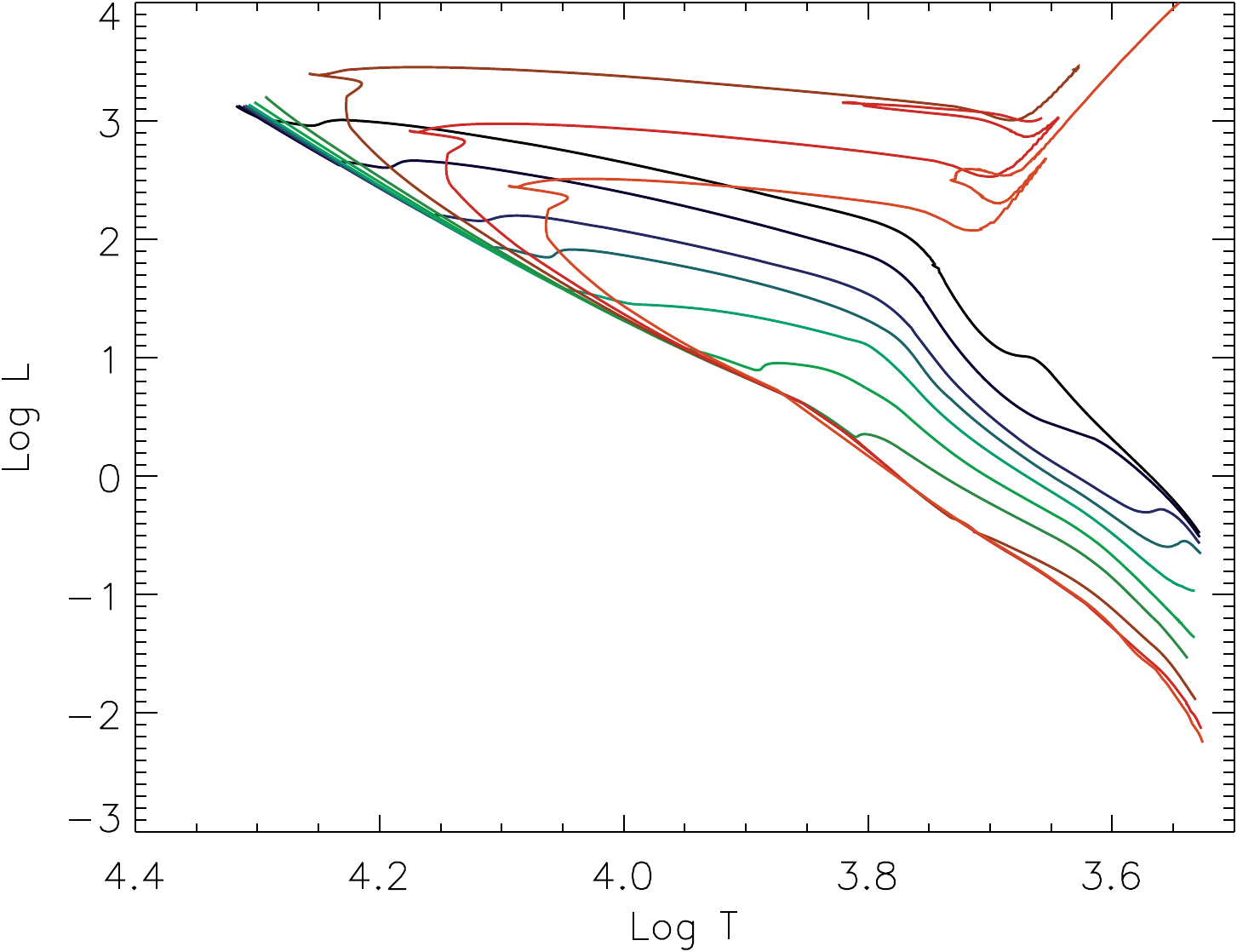}
\caption{A sample of isochrones computed by us using the {\sc FRANEC} code. The shown isochrones correspond to
ages of 0.5, 1, 2, 3, 5, 10, 20, 50, 100 and 200~Myr. \label{fig:plot_theor_isoch}}
\end{figure}

The evolution of light elements (Deuterium, Lithium, Beryllium and Boron)  is explicitly included in this version of {\sc FRANEC} code.
Deuterium burning plays an important role in the very early stages of PMS evolution for stars with $M \lesssim 0.6$~M$_\odot$.
For these stars the energy provided by the D$(p,\gamma)^3$He reaction is able to slow down the contraction in the initial part
of the Hayashi track. This can be seen in Fig.~\ref{fig:plot_theor_isoch}, where  it can be seen
that the low-mass end of the PMS isochrones are bent towards higher luminosities a  for ages younger than $\sim 5$ Myr.

In order to explore the dependance of our findings on the assumed models, the consideration apart from {\sc FRANEC} of additional
families of PMS evolutionary models \citep[e.g.,][]{dantona94, baraffe97, palla99} would be justified. However, the majority of the
available grids of models are not available for low metallically populations like those in the LMC. The grid of
PMS evolutionary models by \citet{siess2000} is the only one that has been computed also for a subsolar value of metallicity
$Z=0.01$ and covers the mass range which is interesting to us, from the hydrogen burning limit of $\sim 0.1$~M$_{\odot}$ to
the intermediate-mass regime with $M$~\lsim~7~M$_{\odot}$. As a consequence, in our subsequent analysis we consider
both {\sc FRANEC} and \citet{siess2000} models. We will refer to the latter simply as the {\sl Siess} models.

\subsection{Synthetic Photometry} \label{section:coloringisochrones}

In order to compare our photometry with the evolutionary models, we convert them from the physical units
($T_{\rm eff}$, $\log L_{\rm bol}$) into absolute magnitudes in the HST/ACS photometric system, by means
of synthetic photometry on a grid of atmosphere models. As discussed thoroughly in Paper~I, the surface gravity
dependence in the computation of optical colors for PMS stars must be accounted for, especially in the low-mass regime.
Since atmosphere models are generally expressed in terms of $T_{\rm eff}$ and $\log g$, synthetic photometry provides
a precise consideration of the color dependence on surface gravity, allowing to isolate one unique spectrum as the best
representative of the stellar parameters for each point in a theoretical isochrone.

The choice of a reliable atmosphere grid is critical for cool stars \citep{darioONCII}, whose spectral energy distributions (SEDs) are dominated by broad molecular bands.
Moreover, these spectral features are gravity dependant, and since the stellar surface gravity varies during PMS contraction, optical colors are age dependant for young stars.
In Paper~I for the derivation of the IMF of
LH~95 we utilized the {\sc NextGen} \citep{nextgen} synthetic spectra to convert the theoretical Siess models into colors and magnitudes,
extended with the \citet{kurucz93} grid for the highest temperatures ($T_{\rm eff}>8000$~K). While our ACS photometry of LH~95
in Paper~I includes PMS stars down to $\sim0.2$~M$_{\odot}$, the detection incompleteness could not be corrected for masses below $\sim0.4$~M$_{\odot}$. Therefore, the
IMF derived in Paper~I was constructed for masses down to $\sim0.4$~M$_{\odot}$. However, in the present study on the age
of LH~95 and the age distribution of its low-mass PMS stars, the whole sample of detected stars down to $\sim$~0.2~M$_\odot$ can
be used. Therefore we want to refine the atmosphere models we adopt to compute intrinsic colors for the very low stellar masses.

\citet{BCAH98} show that {\sc NextGen} models cannot reproduce empirical solar-metallicity $V$ vs $(V-I)$ color-magnitude diagrams for main sequence populations. The problems set in for $(V-I)\gtrsim2$, which corresponds to \teff$\lesssim3600$ K, and can be attributed to  shortcomings in the treatment of molecules. In particular, \citet{allardAMES} illustrate that the main reason for the failure of  {\sc NextGen} spectra in matching the optical properties of late-type stars is the incompleteness of the opacities of TiO lines and water. In that work they present new families of synthetic spectra ({\sc AMES} models) with updated opacities, and discuss the comparison with optical and NIR data, finding a good match with empirical MS dwarf colors from the photometric catalog of \citet{leggett92}.

\citet{darioONCII}, in their analysis of a multi-color optical survey of the Orion Nebula Cluster down to the hydrogen burning
limit, show that, for young, M-type stars: a) the optical colors are bluer than those of MS dwarfs at the same temperature; b)
intrinsic colors obtained from synthetic photometry using \textsc{AMES} spectra  better reproduce the observed colors; c) for late M-type stars the \textsc{AMES} colors are even slightly bluer than the observed sequence, needing an additional correction based on the empirical data. As we show later on, the predicted colors from {\sc NextGen} models are even bluer than the {\sc AMES} ones for young stars. Moreover, the correction to the {\sc AMES} colors proposed in \citet{darioONCII} is critical for temperatures below our detection limit. As a consequence, we are confident that for the analysis in this paper, the{\sc AMES} models are in any case more appropriate than the {\sc NextGen}, and their still present uncertainties do not play a significant role when used on our LH~05 photometry.

Unfortunately the {\sc AMES} library is available only for solar metallicity, and therefore we proceed by performing an approximation
that provides the best estimate of the correct magnitudes and colors for M-type stars with metallicity $Z=0.01$. In conclusion, we perform
synthetic photometry on the evolutionary models considered here by applying the technique presented by \citet{darioONCII} by making
use of three families of synthetic spectra: (a) {\sc NextGen}, for a LMC metallicity of $Z\simeq0.01$, (b) {\sc NextGen} for solar metallicity $Z=0.02$,
and (c) combined {\sc NextGen} for $T_{\rm eff}>5000$~K and {\sc AMES} for $T_{\rm eff}<5000$~K for solar metallicity.
For $T_{\rm eff}>8000$~K all three used grids are extended using \citet{kurucz93} spectra.

\begin{figure*}
\includegraphics[width=\columnwidth]{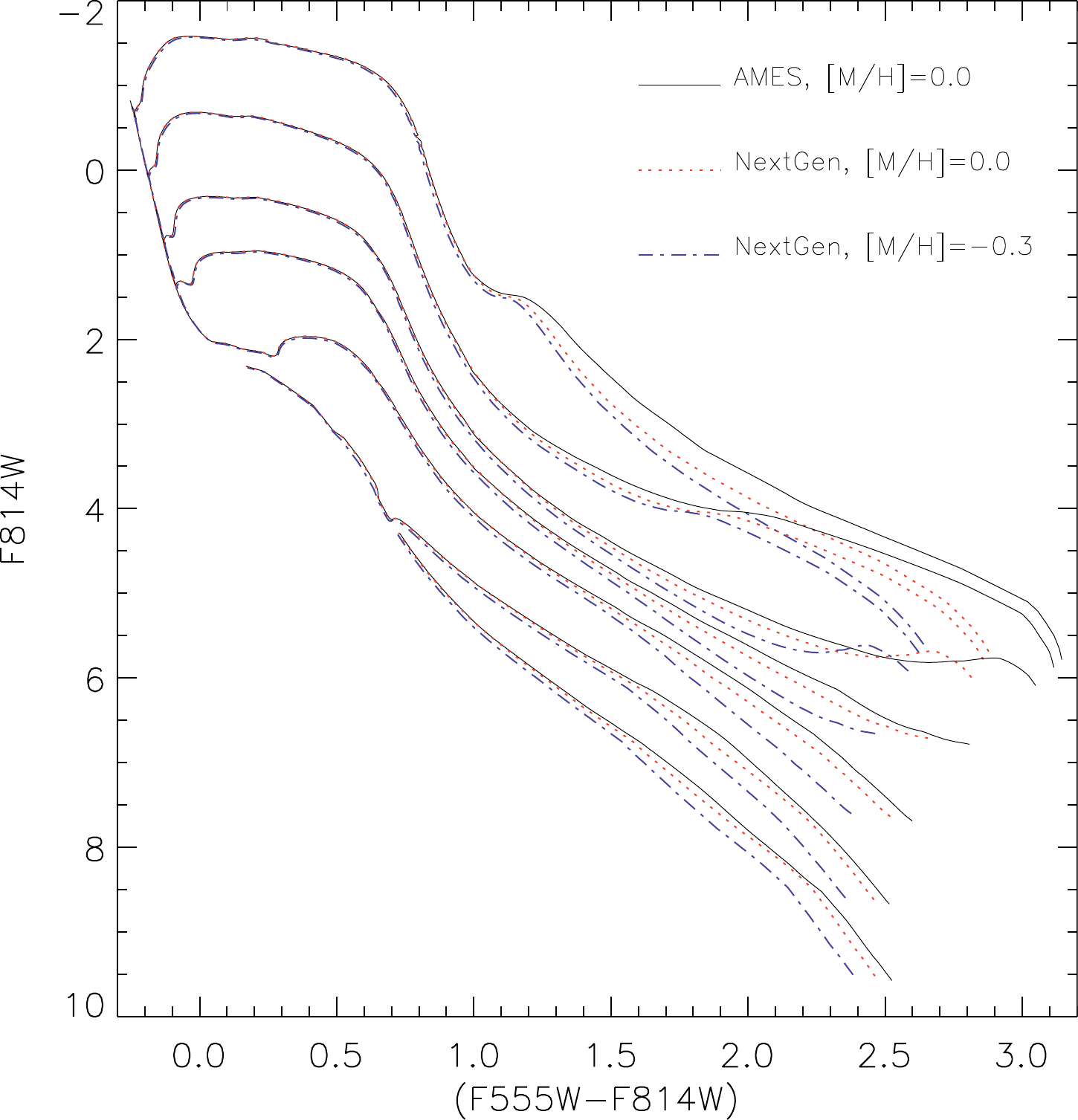}
\includegraphics[width=\columnwidth]{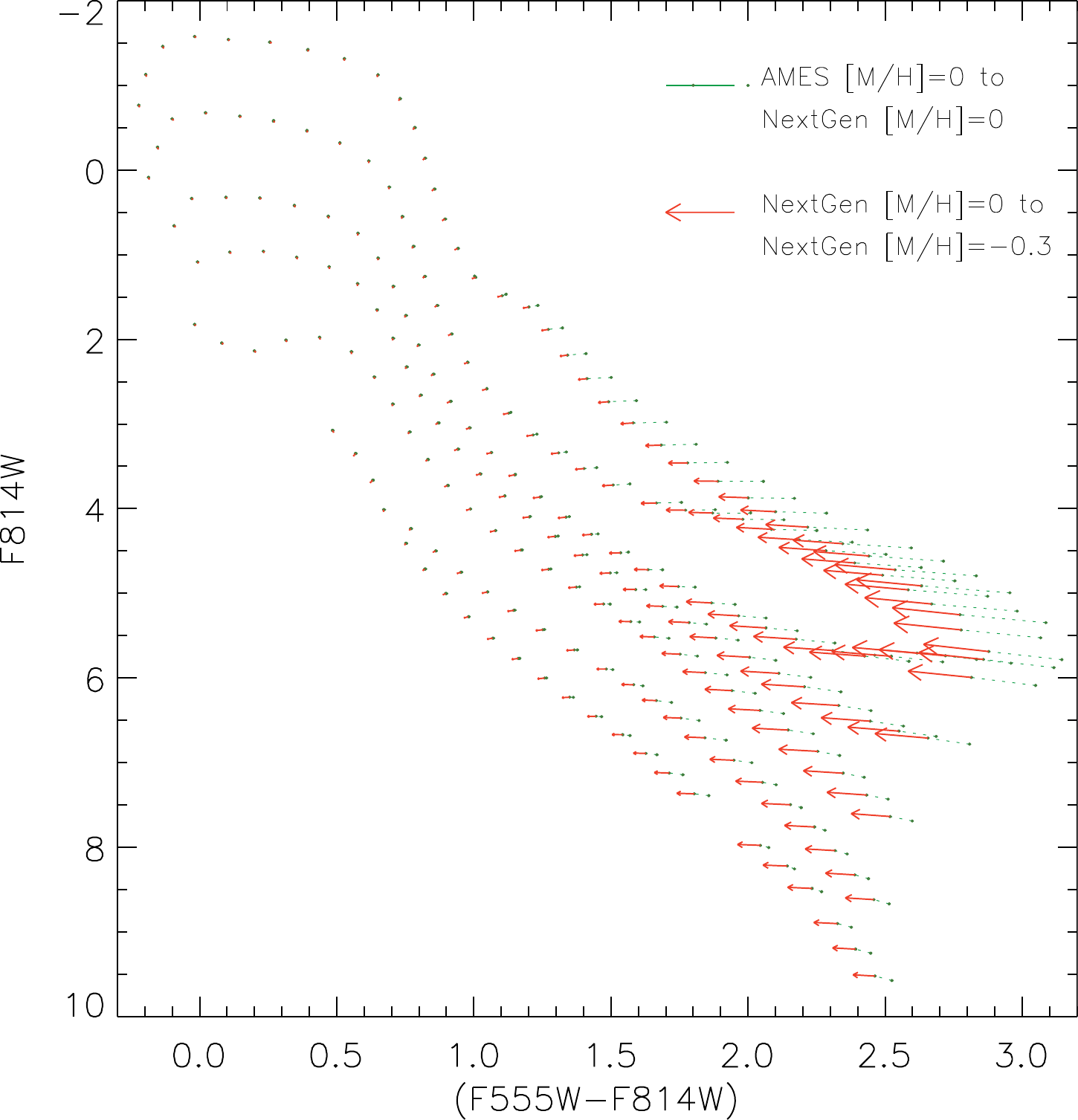}
\caption{{\em Left panel:} CMD showing our {\sc FRANEC} isochrones for ages of  0.5, 1, 3, 5, 10, 30, 100 Myr converted into the ACS
photometric system using the solar metallicity atmosphere models from the {\sc AMES} library (\emph{solid line}), the {\sc NextGen}
models of solar metallicity (\emph{dashed lines}) and the same of LMC metallicity $Z=0.01$ (\emph{dot-dashed lines}). {\em Left panel:}
Representation of the displacements in the CMD of identical points on the evolutionary models caused by changing the three grids of
atmosphere models used. \label{fig:isoch_atm}}
\end{figure*}

Results of our transformations are shown in Fig.~\ref{fig:isoch_atm}({\em left}) for a number of {\sc FRANEC} isochrones with ages spanning
from 0.5 to 100~Myr. In this figure it is shown that while the dependence of colors on the used synthetic spectra is negligible for
low color-terms that correspond to high \teff, the offset increases for late spectral types and young ages. In particular, the more ``correct''
{\sc AMES} atmospheres are up to 0.2~mag redder than the {\sc NextGen} ones for the same metallicity, and a decrease of [M/H] by a factor
of two (from the solar to LMC metallicity) leads to a similar offset towards bluer colors.

In order to estimate the magnitudes and colors of {\sc AMES}-transformed isochrones for the LMC metallicity, we compute the displacement
produced by changing the metallicity from $Z=0.02$ to $Z=0.01$ with the use of the {\sc NextGen} spectra for every point of every isochrone.
These displacements are represented with arrows in Fig.~\ref{fig:isoch_atm}({\em right)}. They can be considered as a fairly good approximation
of the real change in the observed colors and magnitudes in relation to \teff\ and $\log g$, caused by the decrease of metallicity. Therefore, we apply
the same offset, point by point, to isochrones transformed with the use of {\sc AMES} spectra with solar metallicity, in order to derive their counterpart
isochrones for $Z=0.01$.

\begin{figure}
\includegraphics[width=\columnwidth]{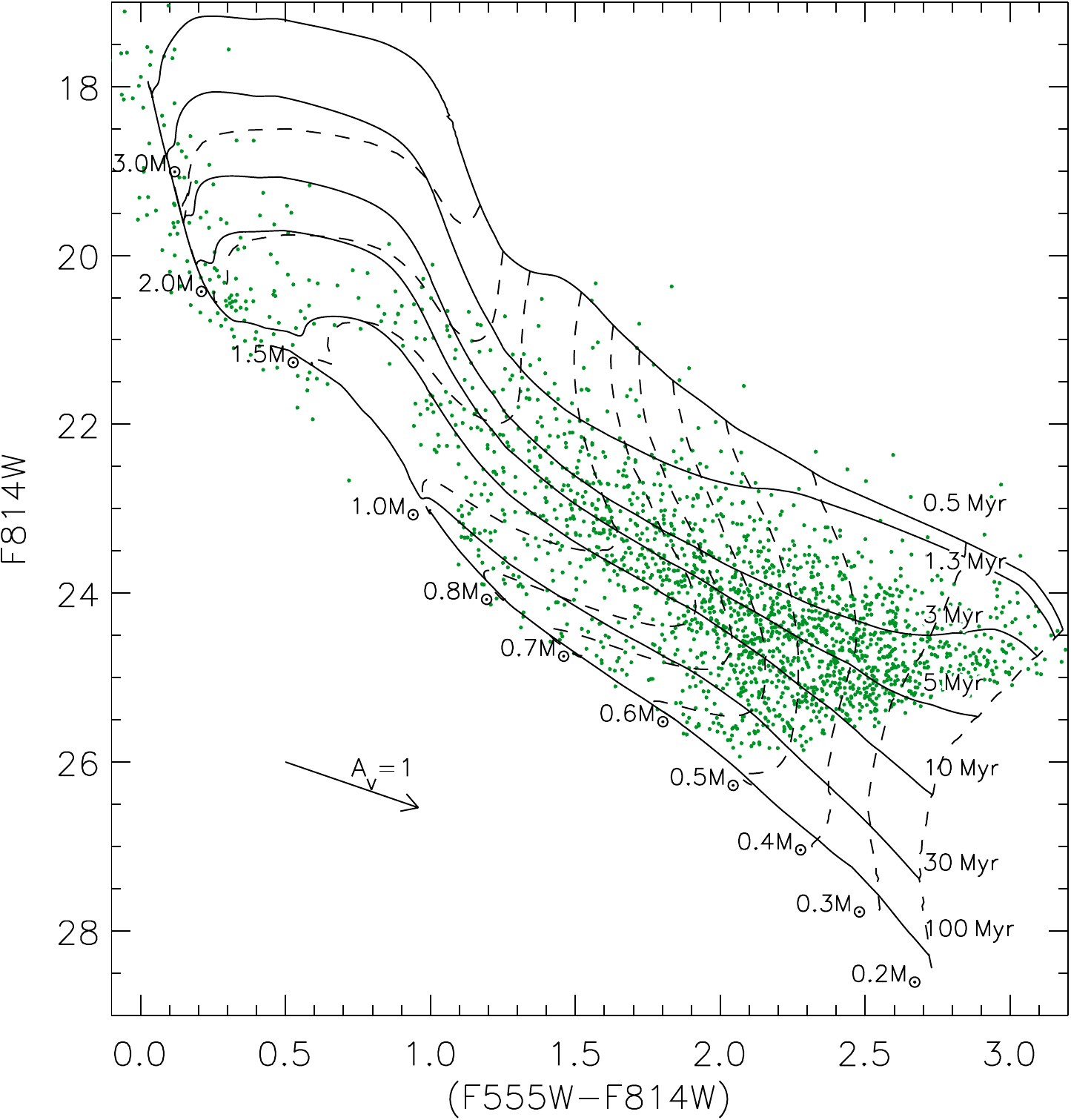}
\caption{The stellar population of LH~95 with, overlaid, the {\sc FRANEC} models converted into observable colors and magnitudes by extrapolating the results
of the {\sc AMES} model atmospheres for $Z=0.01$.  \label{fig:isoch_on_data}}
\end{figure}

Fig.~\ref{fig:isoch_on_data} shows a sample of {\sc Franec} isochrones transformed to the observable plane with the use of the {\sc AMES} spectra
library extrapolated to the LMC metallicity plotted over the CMD of LH~95. A distance modulus of $m-M=18.41$~mag and average extinction of
$A_V=0.6$~mag, derived in Paper~I, are applied to the isochrones. The PMS stars of LH~95, roughly following the 5~Myr isochrone, exhibit a
wide apparent spread that spans from the youngest ages available in the models to the main sequence.

\begin{figure*}
\includegraphics[width=\columnwidth]{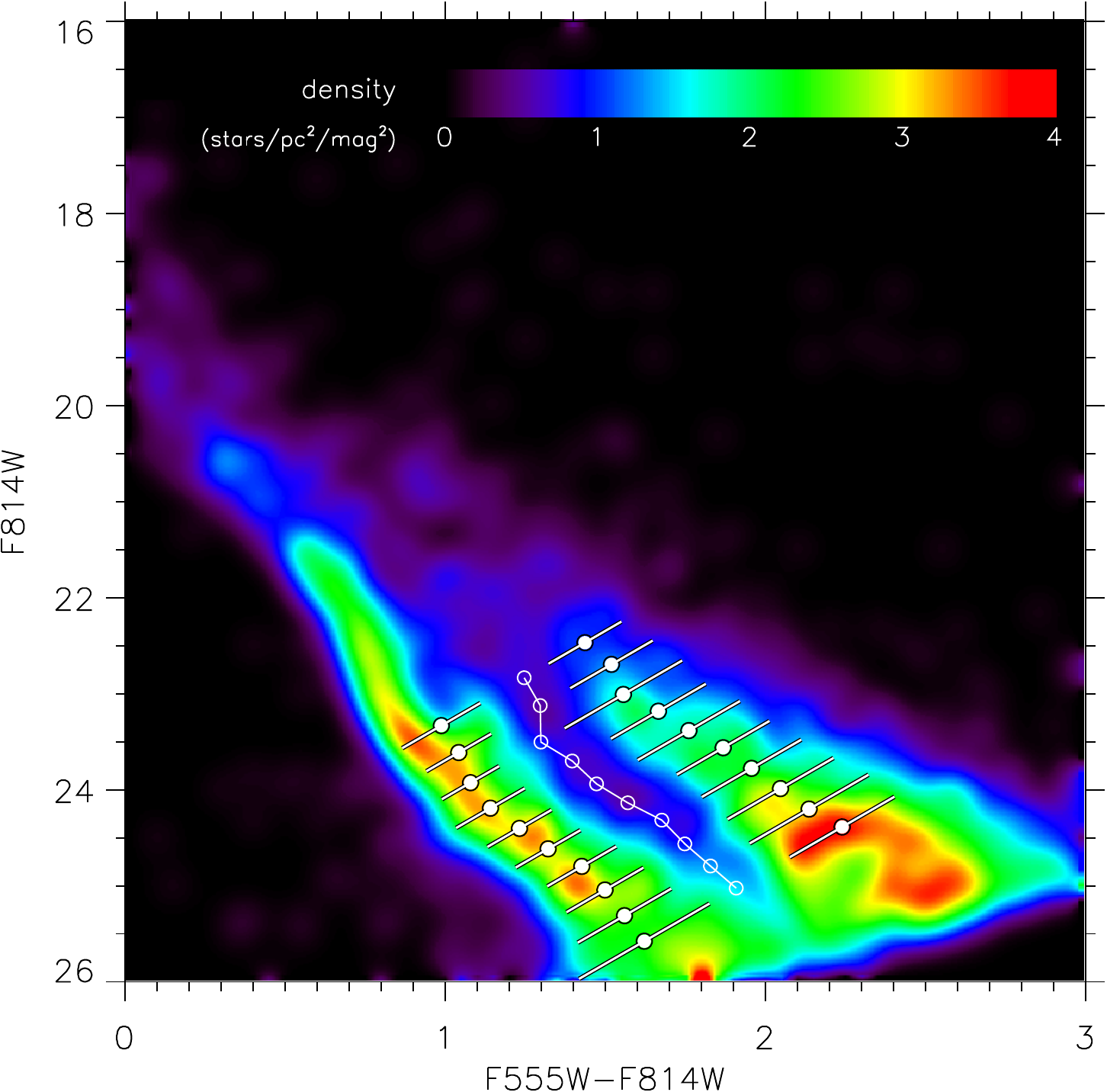}
\includegraphics[width=\columnwidth]{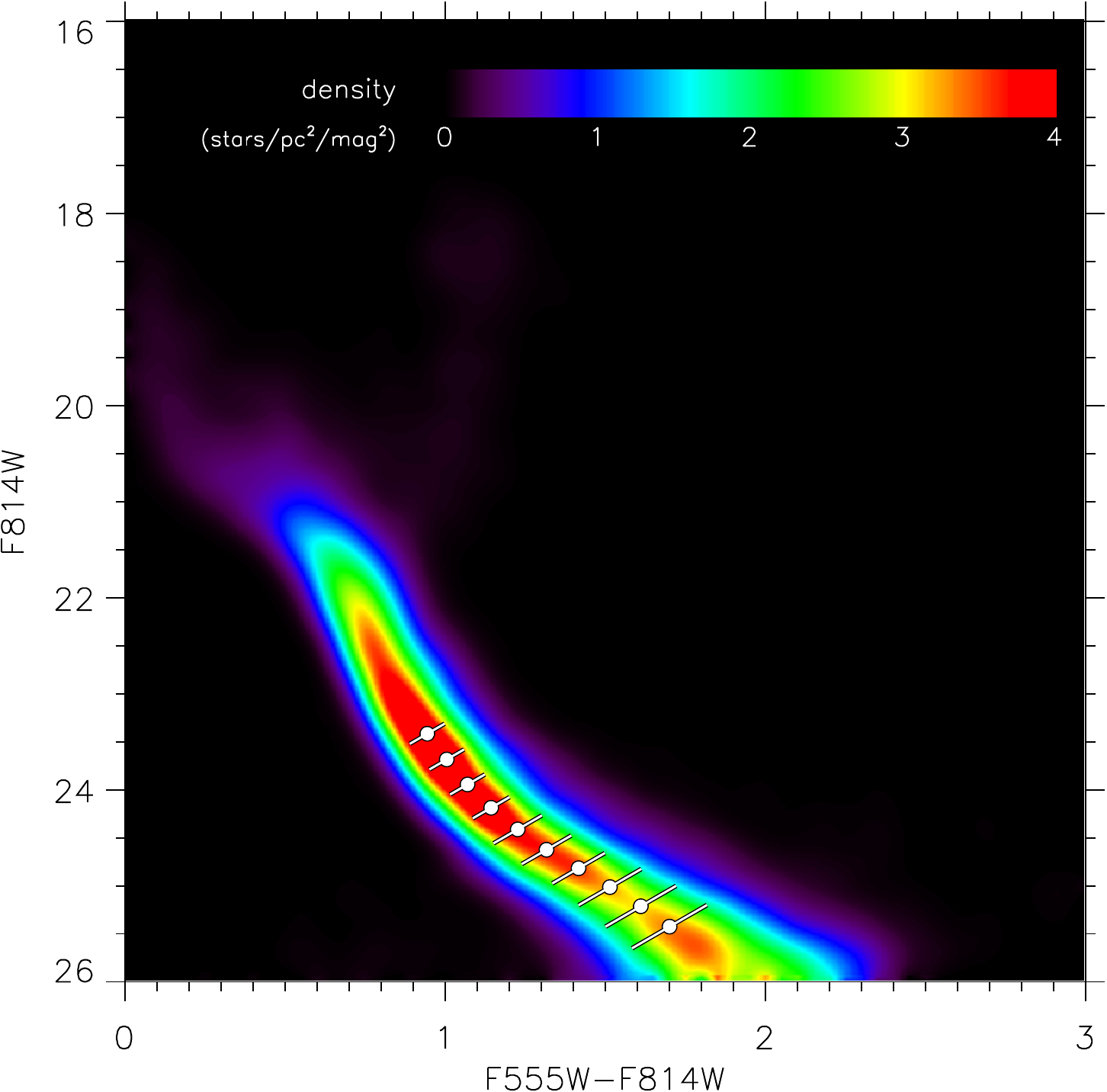}
\caption{{\em Left}: Hess diagram of the stellar density in the area of LH~95, with no field subtraction. There are two distinct stellar sequences
apparent, the MS attributed to the LMC field and the native PMS population. The results of the simultaneous fit of two Gaussians along diagonal
cross-sections are drawn, using filled circles to indicate the centers of the Gaussians for every strip, open circles for the minima between them,  and
white lines following the directions of the strips for the standard deviations of the Gaussians. {\em Right}: The same as in the left panel for the area
of the control field. This CMD demonstrates the complete lack of any PMS population in the local LMC field, making the distinction of the true PMS
stellar-members of LH~95 quite straightforward. A single Gaussian is used to fit the MS of this CMD along identical cross-sections as for LH~95.
\label{fig:CMD_trimming}}
\end{figure*}

\section{Field subtraction refinement} \label{section:fieldsub_refinement}

Our observations of the control field 2\arcmin\ away from LH~95 with setup identical to those of the system allowed us in Paper~I
to perform a thorough statistical removal of the contamination of the observed population by the stars of the local LMC field, on the
verified basis that there are no PMS stars in the LMC field. Indeed, this method provided a sufficiently clean CMD of the PMS
cluster-members of LH~95 (shown in Fig.~\ref{fig:isoch_on_data}). However, there is an anomalous presence of a small
number of sources in the field-subtracted CMD located between the general PMS population and the MS. These sources can
be seen in Fig.~\ref{fig:isoch_on_data} located in the neighborhood  of the 100 and 30~Myr isochrones. In Paper~I we hypothesized
the origin of these contaminants as  a binary sequence of the field MS population, and we demonstrated that their removal does not affect
the shape of the derived IMF.

However, for the present study, which focuses on the identification of any age distribution among PMS stars, the nature of these
sources and their membership should be reevaluated. In order to study the spread of the field MS in the CMD of both LH~95 and
the control field, we construct the Hess diagram of the complete stellar sample observed in the area of the association before any
field subtraction. We bin the stars in the color- and magnitude-axis of the CMD and smooth the counts with a Gaussian kernel of
0.05~mag in $m_{\rm 555}-m_{\rm 814}$ and 0.16~mag in $m_{\rm 814}$. We repeat this process to construct the Hess diagram
of the control field. The constructed Hess diagrams are shown in Fig.~\ref{fig:CMD_trimming}. An important difference between the
field population covered in the area of LH~95 and that in the control field is that LH~95 is embedded in the {\sc Hii} region
named LHA~120-N~64C \citep[or in short N~64C,][]{henize56} or  DEM~L~252 \citep{davies76}, and therefore the reddening of
the background field stars in the area of the system should be somewhat higher than those in the control field. However, a comparison
of the Hess diagrams of Fig.~\ref{fig:CMD_trimming} shows that there is no significant variation of the field population between the two
areas, and therefore the appearance of the nebula does not seem to affect significantly the field MS. This suggests that the LMC field population is
mainly distributed in foreground with respect to LH95. Nevertheless, a small fraction of field stars could be located behind the nebula, and be affected by a larger extinction.

\begin{figure}[t!]
\includegraphics[width=\columnwidth]{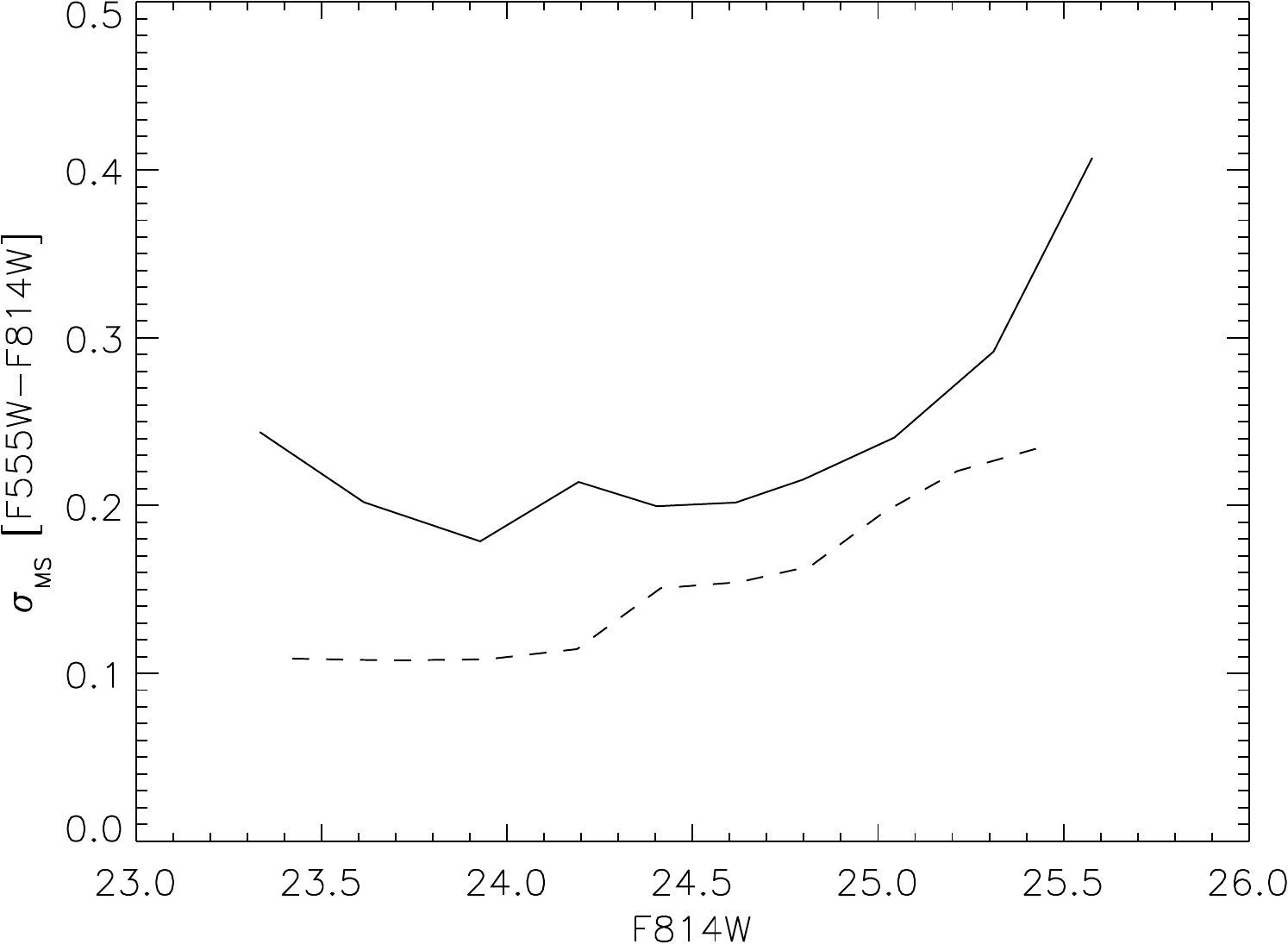}
\caption{CMD broadening of the MS in color as a function of luminosity for the area of LH~95 (solid line) and the control field (dashed line). The
smaller broadening of the MS in the control field than in LH~95 demonstrates that field stars could not be completely removed within the application
of the statistical field-subtraction in Paper~I. Here we refine our selection of true stellar-members for LH~95, based on the results of Gaussian fits to
the distribution of the MS in the field and the system. \label{fig:CMD_trimming_sigmadiff}}
\end{figure}

\begin{figure}
\includegraphics[width=\columnwidth]{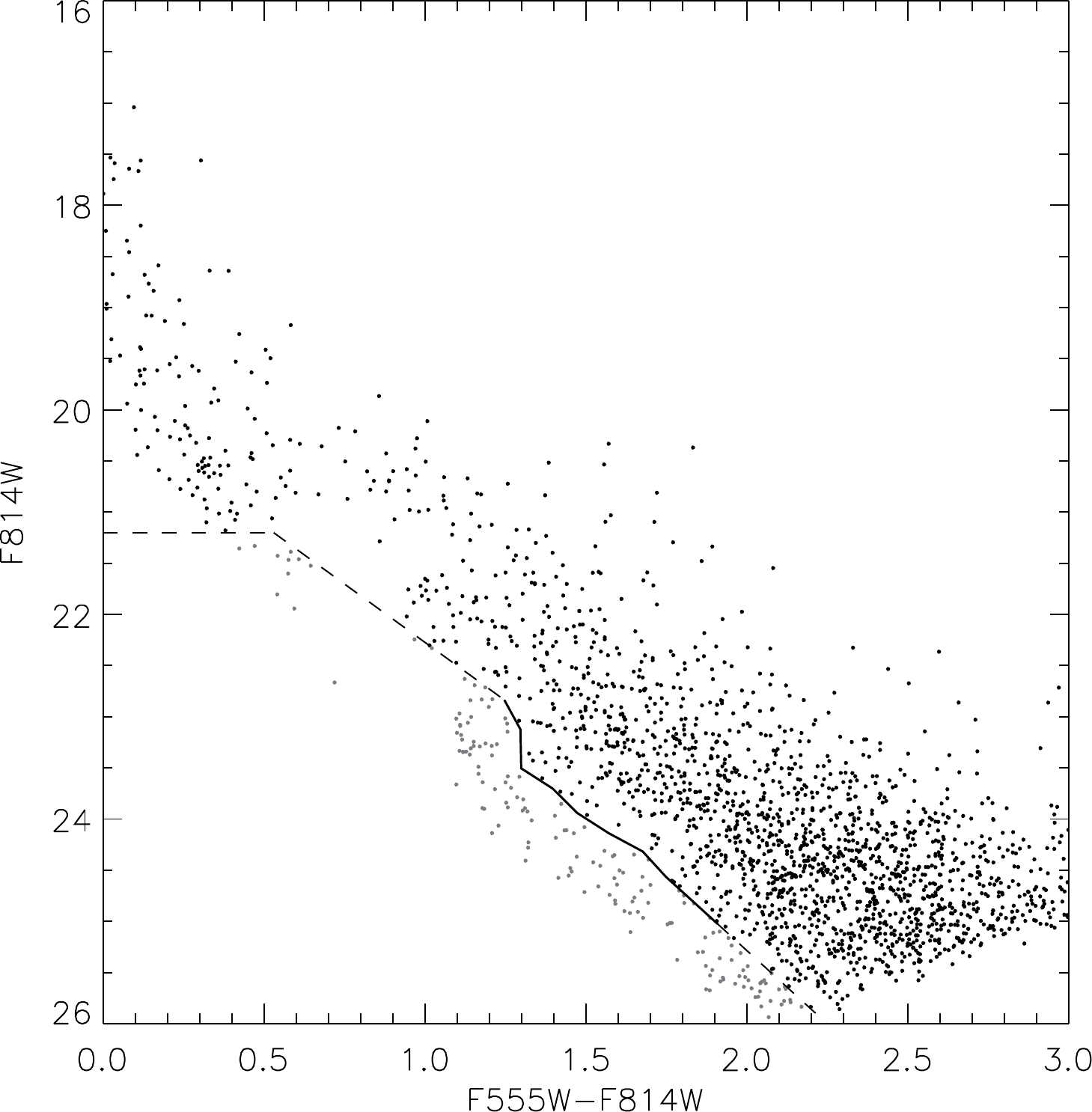}
\caption{Refinement of the removal of the field contamination in the CMD of LH~95. The drawn solid line follows the minima of the best-fitting
Gaussians to the MS and PMS stellar distributions before any field subtraction (see Fig.~\ref{fig:CMD_trimming}{\em left}), showing that the
bluest sources of the population clearly belong to the MS. The dashed line is our chosen continuation of the previous, at higher and lower luminosities. Sources located to the blue of this MS border and its extensions (dashed lines)
are excluded from our analysis as true field-members. \label{fig:CMD_trimming2}}
\end{figure}

In order to test this possibility and further eliminate the sample of PMS cluster-members from potential field MS contaminants, we consider diagonal cross-sections
of width 0.33~mag in the direction defined by the line $m_{\rm 814}=-1.92\times(m_{\rm 555}-m_{\rm 814})$, and we fit with Gaussians
the distributions of the stars along each of these strips. A sum of two Gaussian distributions is considered for the fit of the CMD of LH~95
and a single Gaussian for that of the control field. The centers and standard deviations, $\sigma$, of the best-fitting Gaussians are drawn
over the Hess diagrams of Fig.~\ref{fig:CMD_trimming}. We compare the $\sigma$ of the best-fitting Gaussians to the MS, as a measurement
of the broadening of the MS, between the CMD of LH~95 and that of the control field in Fig.~\ref{fig:CMD_trimming_sigmadiff}, where we
plot $\sigma$ for each selected strip as a function of the corresponding average luminosity. In this figure can be seen that $\sigma$
and therefore the MS broadening is systematically smaller in the control field than in LH~95, in agreement to what we discussed above.
This difference demonstrates quantitatively that the contamination of the CMD of LH~95 by field stars cannot be 100\% removed, and this
applies mostly to the reddest part of the field MS in the area of LH~95 due to higher reddening. As a consequence we further remove from
the sample of cluster-member sources those that fall blue-ward of the minima between the two gaussian fits of the CMD of LH~95, shown in
Fig.~\ref{fig:CMD_trimming}({\em left}). This second selection border is drawn on the CMD of the area of LH~95 in Fig.~\ref{fig:CMD_trimming2}; we prolong this border at lower and higher luminosities, as shown with a dashed line in Figure \ref{fig:CMD_trimming2}. At the faint end a linear extrapolation in the CMD is chosen; for the bright end we arbitrarily define a cut in order to remove a handful of stars at $I\sim22$~mag, undoubtedly located on the MS.
There are 162 stars, which we consider to be field subtraction remnants, plotted with gray dots to the left of the MS border in
Fig.~\ref{fig:CMD_trimming2}. These sources will be excluded from our subsequent analysis.

We stress that, despite the improvement of this field subtraction refinement with respect to that of Paper~I, some uncertainty in the membership assignment on the tail of the distributions (close to the cut shown in Figure \ref{fig:CMD_trimming2}) are still present. This effect, which might affect the distribution of the oldest stars, should not be significant. The population, within the central region, considered MS field is about as numerous as that considered part of the young system (1904 and 1942 stars respectively). As a consequence, the number of young members erroneously removed being too blue should be roughly compensated by a similar number of MS stars redder than the minima between the two Gaussian (Figure \ref{fig:CMD_trimming}). This guarantees that the number of old PMS members of stars is fairly accurate, but could lead to an underestimation of the ages of some of those. In any case, since in this work we demonstrate a significant age spread for LH95, this fact would just reinforce our findings.

\section{Age determination of PMS stars}
\label{section:age_estimation_main}

In this section we apply our new maximum-likelihood technique for the derivation of the ages of the PMS stars of LH~95 from
their photometry. This method allows the determination of the age of a PMS star from its position in the CMD in a probabilistic manner,
accounting for the physical and observational effects that affect the colors and magnitudes of all observed PMS stars.

In order to estimate the age of a young stellar system when only photometry is available, it is common practice to simply overlay
a set of isochrones to the CMD and, by interpolation, assign an age to all the members. This approach, however, can lead to
inaccurate results \citep{gouliermis-eslab07}. This is because every star is scattered from its original position in the CMD by a
number of factors, and specifically by photometric errors, differential reddening, unresolved binarity, stellar variability,
non-photospheric excess due to accretion, and scattered light from circumstellar material.

These effects, when not accounted for, bias the observed age distribution, both increasing the observed spread and systematically
shifting the mean cluster age. These factors can be disentangled {\sl only} on the basis of individual stars, which is the case when
spectroscopic measurements are available for all stars. However, such observations on large numbers of stars are not technically
feasible, and photometric observations are the most appropriate for collecting data on vast stellar numbers, like in the case of LH~95.
In this case, a convenient way to estimate the overall effect of observational and physical biases to the original CMD positions of PMS
stars is to compare the observed data with synthetic CMDs, which are constructed by the application of these biases to one or more
theoretical PMS isochrones. There are some examples of such an approach presented in the literature, especially for the
investigation of the relation between the observed luminosity spread and the true age-spread among PMS star-members of young
stellar systems \citep[e.g., ][]{burningham2005,jeffries2007_2169,hennekemper2008,hillenbrand2008}.

In this context, \citet{naylor-jeffries06} (hereafter NJ06) presented a rigorous approach to derive the age of young stellar systems accounting
for photometric errors and binarity. They present a maximum-likelihood method for fitting two-dimensional
model distributions to observed stellar CMDs. In their application, the two-dimensional models are probability density distributions in the CMD,
obtained by applying binarity to a PMS isochrone of a given age, considering that the model representing a coeval population in the CMD is not
a line -- like a theoretical isochrone -- but a 2D density distribution. There are several advantages in the application of this approach. First, it
enables one to account rigorously for the photometric errors in both color and magnitude when the fit is performed. Second, binarity is considered
in the fitting process itself. Finally, this method allows the evaluation of its goodness-of-fit.

In the present study we further develop the method of NJ06, by extending its application to a realistic case of a PMS population, including in the
computation of the two-dimensional models not only binarity, but also variability, crowding, and differential extinction.

\subsection{From isochrones to density distributions in the CMD}
\label{section:2Disochrones}

As in NJ06, we wish to construct a stellar density distribution in the CMD for every age in the considered set of
evolutionary models. This distribution represents the probability for a star of an assumed age to be found at
a given point in the CMD. We consider the two families of evolutionary models ({\sc FRANEC} and {\sl Siess}) that
we converted into magnitudes in the HST/ACS photometric system in \S~\ref{section:coloringisochrones}.
For every considered age, we populate the appropriate isochrone with $5\cdot10^6$ simulated stars. We assume
the IMF of LH~95 derived in Paper~I and correct it for binarity. The change in the measured  IMF slopes caused by
 binarity was already investigated in Paper~I, through a Monte-Carlo simulation. Here we adopt these results,
 summarized in Table \ref{table:IMFslopes}. We found that the IMF for intermediate and high-mass stars ($M>1$~M$_\odot$) is almost insensitive on binaries, whereas for low-mass stars, the slope changes of up to $0.4$ units adding an increasing fraction of unresolved binaries. Neglecting binaries leads to a shallower measured IMF.

\begin{deluxetable}{lrr}
\tablecaption{IMF slopes corrected for binarity, as a function of binary fraction}
\tablehead{
\colhead{$f$} &
\colhead{$x$ $(M<1M_\odot)$}  &
\colhead{$x$ $(M\geq1M_\odot)$} \\
}
\startdata
   0.0  \tablenotemark{$\dagger$} &     1.05 &  2.05  \\
   0.2 &     1.20 &  2.05  \\
   0.4 &     1.30 &  2.05  \\
   0.5 &     1.35 &  2.1  \\
   0.6 &     1.40 &  2.1  \\
   0.8 &     1.45 &  2.1
\enddata
\tablenotetext{$\dagger$}{Measured IMF from Paper~I.}
\tablecomments{In the units we use to express IMF slopes, a \citet{salpeter55} IMF has a slope $x=1.35$. }
\label{table:IMFslopes}
\end{deluxetable}

For smaller masses, not having a direct measurement of the IMF in LH~95, we assume
a value $x=0.3$ compatible with the \citet{kroupa2001} IMF. This value was also assumed in Paper~I to compute the binarity-corrected IMF slopes reported in Table \ref{table:IMFslopes}.
It is worth noting that small changes in the slope of the assumed IMF
do not influence significantly the ages for single stars derived using our method (see Section \ref{section:deriving_single_ages}), but could affect the evaluation of the goodness-of-fit (Section \ref{section:deriving_age_spread}) we use to determine the age spread. 

After simulating a population of stars for every age, (for the two families of evolutionary models, and all the binary fractions $f$ reported in Table \ref{table:IMFslopes},
we displace with a random process the stars in the CMD according to the dictations of several
factors. They are discussed below.

\begin{itemize}

\item[{\em (i)}] {\em Rotational variability:} We simulate statistically the variation of  magnitudes in the F555W and F814W bands due to variability caused
by stellar rotation. We consider the results of \citet{herbst2002}. These authors surveyed the PMS population of the Orion Nebula
Cluster (ONC) in the $I$-band over 45 nights, obtaining the light curves for $\sim1500$ candidate members. We consider their catalog
of peak-to-peak variations. A single imaging of a young stellar system, like LH~95, statistically measures the flux of variable stars in
random phases of their variation. Considering, at this point, variability only due to dark spots on the stellar surface, the measured fluxes are
equal or fainter than those predicted if no variability is considered. Therefore, if we assume for simplicity sinusoidal light curves, the
statistical distribution of $I$-band variations for a star with a measure peak-to-peak $\Delta I_{\rm max}$ is the distribution of
$\Delta I_{\rm max}\cdot(\sin x +1)/2$ with $0<x<2\pi$. We compute this for every star of the catalog of \citet{herbst2002} and
combine the derived distributions. The result, shown in Fig.~\ref{fig:variab_deltaI_distrib}a, is the observed $I$-band variation
distribution that we assume as valid for a single epoch photometry in a young stellar system. The peak at very low $\Delta I$ is mainly due to the large fraction of stars not showing significant rotational variability. We displace the brightness in
the F814W filter  of each of the simulated stars along each isochrone by a value randomly drawn from this distribution.

The amplitude of the $V-$band variation due to dark spots is generally larger than in $I$, and the ratio between the two depends, in general, on the temperature of the star-spot, on the temperature of the photosphere, and the filling factor of the spot. According to the results of \citet{herbst1994}, the ratio $\alpha = \Delta I/\Delta V$ spans from 0.39 to 0.88. The variations in the two bands should be correlated if the observations were close to co-temporal. In our photometry, however, as mentioned in Section \ref{s:dataset}, $V$ and $I$ observations were taken separately, one 4 days after the other. According to \citet{herbst2002}, the rotational period distribution of T-Tauri stars is bimodal, with a peak at $P\sim2$~days and another at $P\sim8$~days. Even considering the longest periods, the time delay of our observations covers about half period. As a consequence, we cannot consider the $V-$ and $I-$band variations to be correlated. We therefore simulated the $V-$band variation distribution from the $I-$band one, multiplying each of the simulated $I$ values for a factor $1/\alpha$, where $\alpha$ was randomly chosen within the range 0.39-0.88. Then these $V$ variations were applied randomly (regardless $\Delta I$) to all the simulated stars.

\item[{\em (ii)}] {\em Accretion:} On-going accretion from the circumstellar disk onto the stars also affects the observed fluxes. In the optical wavelength range, two components contribute to the accretion excess: an optically thin emission generated in the infalling flow, and an optically thick emission from the heated photosphere below the shock \citep{calvetgullbring98}. The optically thin emission is characterized by a line-dominated spectrum, similar to that of a HII region; in particular the bright Balmer emission is commonly used to derive mass accretion rates for PMS stars \citep[e.g.,][]{demarchi2010}. In $V-$ and $I-$ band the optically thick emission dominates, producing an excess, in spectroscopy reffered to as ``veiling'', analogous to the presence of a hot spot on the stellar photosphere. For a given accreting star, the spectrum and intensity of the excesses may vary with time, due to both variations in the mass accretion, and stellar rotation.

    In order to model the overall effect of accretion on a large sample of low-mass stars, we utilize the results of \citet{darioONCII}. In this work the stellar population of the Orion Nebula Cluster was studied from multi-band optical photometry and spectroscopy. In particular, for every star, the veiling excesses were determined from the comparison of the observed colors with the intrinsic ones, after modeling the color displacements caused by accretion and extinction. For the ONC, the authors find about 60\% of the members showing evidences of accretion veiling. We consider this stellar sample, and restrict to the mass range $0.3$M$_\odot\leq M\leq 3$M$_\odot$, relevant to our study. Then, we displace each of the simulated stars for every isochrone of an amount $\Delta V$ and $\Delta I$ equal to the veiling of a randomly chosen Orion member from \citet{darioONCII}. An example of the distribution of displacements in color and magnitude is shown in Figure \ref{fig:variab_deltaI_distrib}b.

    We stress that, statistically, the effect of variability in the mass accretion is already included in the results of \citet{darioONCII}; this is because their observation produced a ``snapshot'' of the veiling excesses for $\sim 1000$ members at a given time.
    It could be argued that the members of LH~95, with an average age about twice as older as Orion, could be affected by less veiling than for the Galactic cluster, since mass accretion decreases with stellar age  \citep{sicilia-aguilar2006}. With no direct measurements of accretion in LH~95 we are unable to accurately constrain the real distribution of veiling excesses for our region. Nevertheless, there are evidences \citep[e.g.,][]{romaniello2004,demarchi2010} suggesting that accretion lasts longer in the Magellanic Clouds, because of the lower metallicity. Because of this we are confident that the distribution of $V$ and $I$ excesses we assumed could be appropriate for LH~95. Future measurements of mass accretion in our region (e.g., from $H\alpha$ photometry) could help to better constrain this parameter.

\item[{\em (iii)}] {\em Binarity:} For every isochrone, we associate pairs of stars from the simulated sample. This is done randomly, but allowing only configurations within a mass range, defined by constraining
the mass ratio to be a flat distribution from 0.2 to 1 \citep{woitas2001}, and do not allow binary systems to form with mass ratio $<0.2$.
We let the binary fraction be a free parameter, and compute different sets of two-dimensional models for different choices of its value.
This is because, as we discuss later in detail, the assumed binary fraction influences significantly the derived ages, and therefore we
present and compare findings obtained with different choices of this parameter. We assume a unique binary fraction and mass ratio distribution
for the whole PMS mass-range. In principle high-mass stars are known to have both higher binary fraction and higher binary mass ratio than
intermediate and low-mass stars \citep[e.g.,][]{lada06,zinneckerARAA07}. However, this does not affect our treatment, since in our analysis
of stellar ages we only consider PMS stars and not the `evolved' high-mass MS stars of LH~95.

\begin{figure}[t!]
\includegraphics[width=\columnwidth]{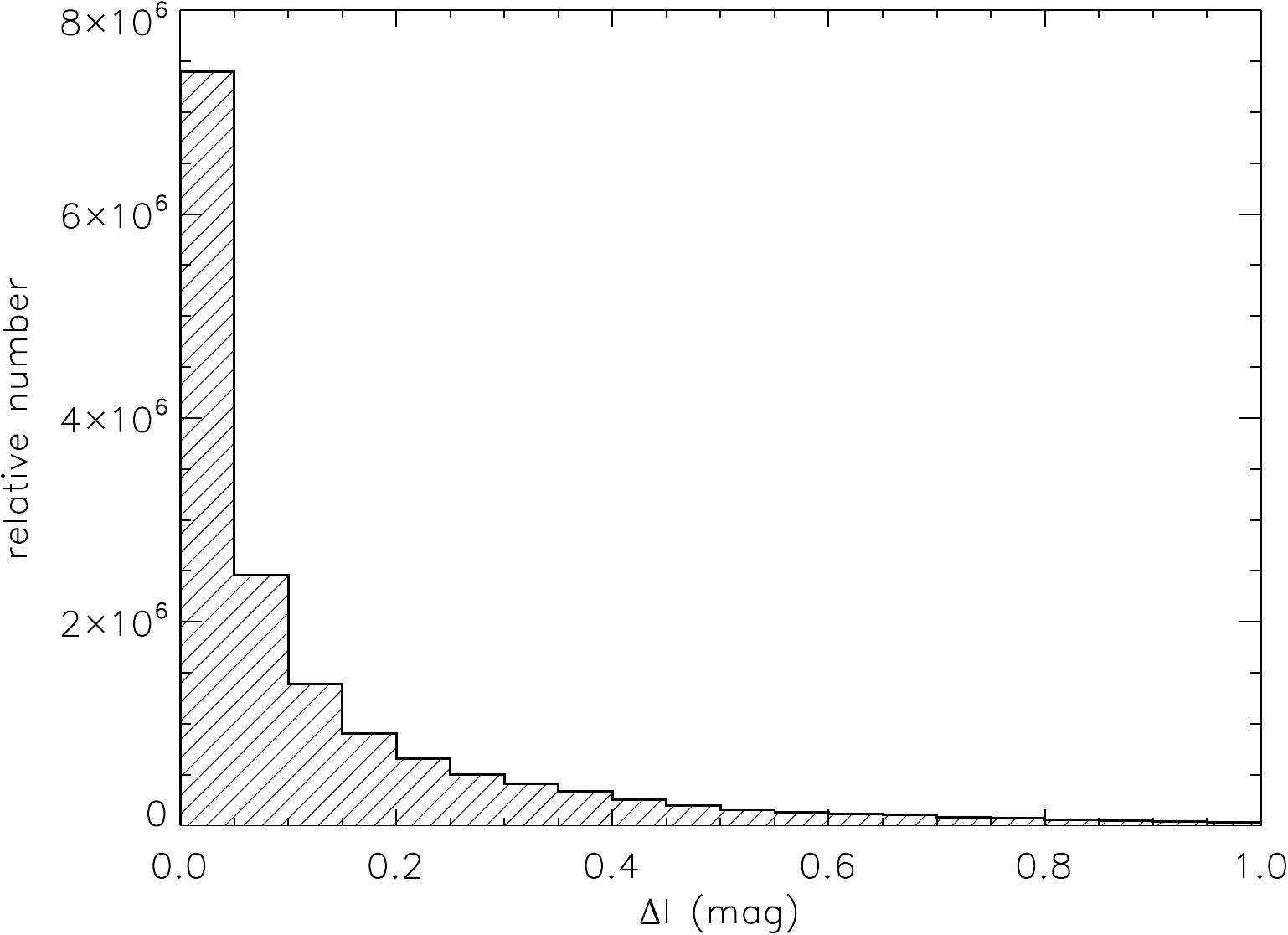}
\includegraphics[width=\columnwidth]{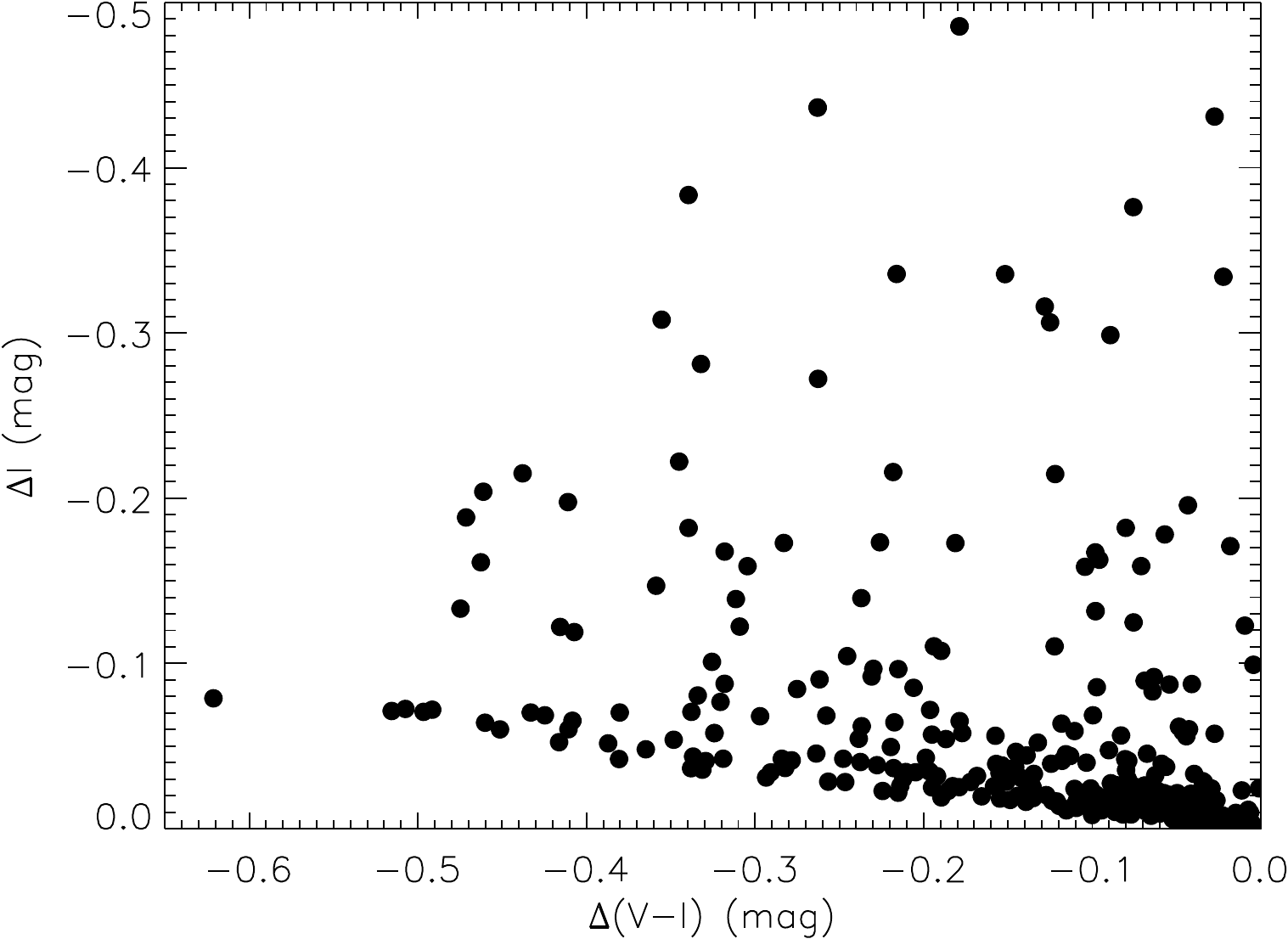}
    \caption{{\em Upper panel: }Distribution of the expected $m_{\rm 814}-$band variation due to dark spots and stellar rotation in single epoch photometry,
    obtained from the peak-to-peak distribution of \citet{herbst2002}. {\em Lower panel: }Displacements in color and magnitude caused by accretion veiling, from the findings of \citet{darioONCII}. \label{fig:variab_deltaI_distrib}}
\end{figure}

\item[{\em (iv)}]  {\em Differential extinction:} In Paper~I we derived the extinction distribution for LH~95, measuring the reddening of the upper
main-sequence (UMS) from the ZAMS. In that work not all the UMS were considered to derive the $A_V$ distribution, but some of the reddest were removed, since we suspected them to be Herbig Ae/Be stars.
The average extinction is quite low, equal to $A_{\rm 555} \simeq0.6$~mag, and the its distribution spans from
$A_{\rm 555} =0$ to $A_{\rm 555} \simeq1$. We use the relation $A_{\rm 555} /A_{\rm 814} =1.85$, as it was computed in Paper~I for the ACS
photometric system, using the extinction law of \citet{cardelli}. In our simulations we apply a random value of extinction to each star, drawn
from the real measured $A_{\rm 555} $ distribution of LH~95.

\item[{\em (v)}]  {\em Source confusion:} LH~95 is a loose association, not centrally concentrated, which comprises at least 3 subclusters of somewhat
higher stellar densities. As a consequence, source confusion due to crowing does not play a dominant role. Nevertheless, we include
in our simulations a 5\% source confusion, by adding the flux of this fraction to sources regardless of their assumed binarity, i.e., by
allowing for random confusion also between binaries and between binaries and isolated stars. We stress that, since this additional pairing is done randomly, because of the IMF the modeled source confusion is also magnitude dependant, favoring the disappearance of faint systems.
\end{itemize}

For the computation of the 2D stellar densities we do not include photometric errors as additional sources of spread.
While the aforementioned biases, which we include in the simulations, can be considered as part of {\em the model}, the photometric
errors are part of {\em the data}, which the model is fitted to, and therefore photometric uncertainty will be consider later in the fitting process (\S~\ref{section:deriving_single_ages}).

It should be finally noted that the constructed 2D distributions are not re-normalized to, e.g., a total integral of 1, because
isochrones
of different ages have different maximum mass. Instead, we use the same sample of 5 million simulated stars with masses drawn
from the assumed IMF and we consider only the fraction included within the limits of each isochrone. In this manner, although the
total number of synthetic stars changes slightly among different 2D distributions, the number of stars per mass interval is constant.
In other words, since our method to assign stellar ages and determine the cluster age is based on relative comparison of probabilities,
an overall normalization of all the models is completely unnecessary.

\begin{figure}[t!]
\includegraphics[width=\columnwidth]{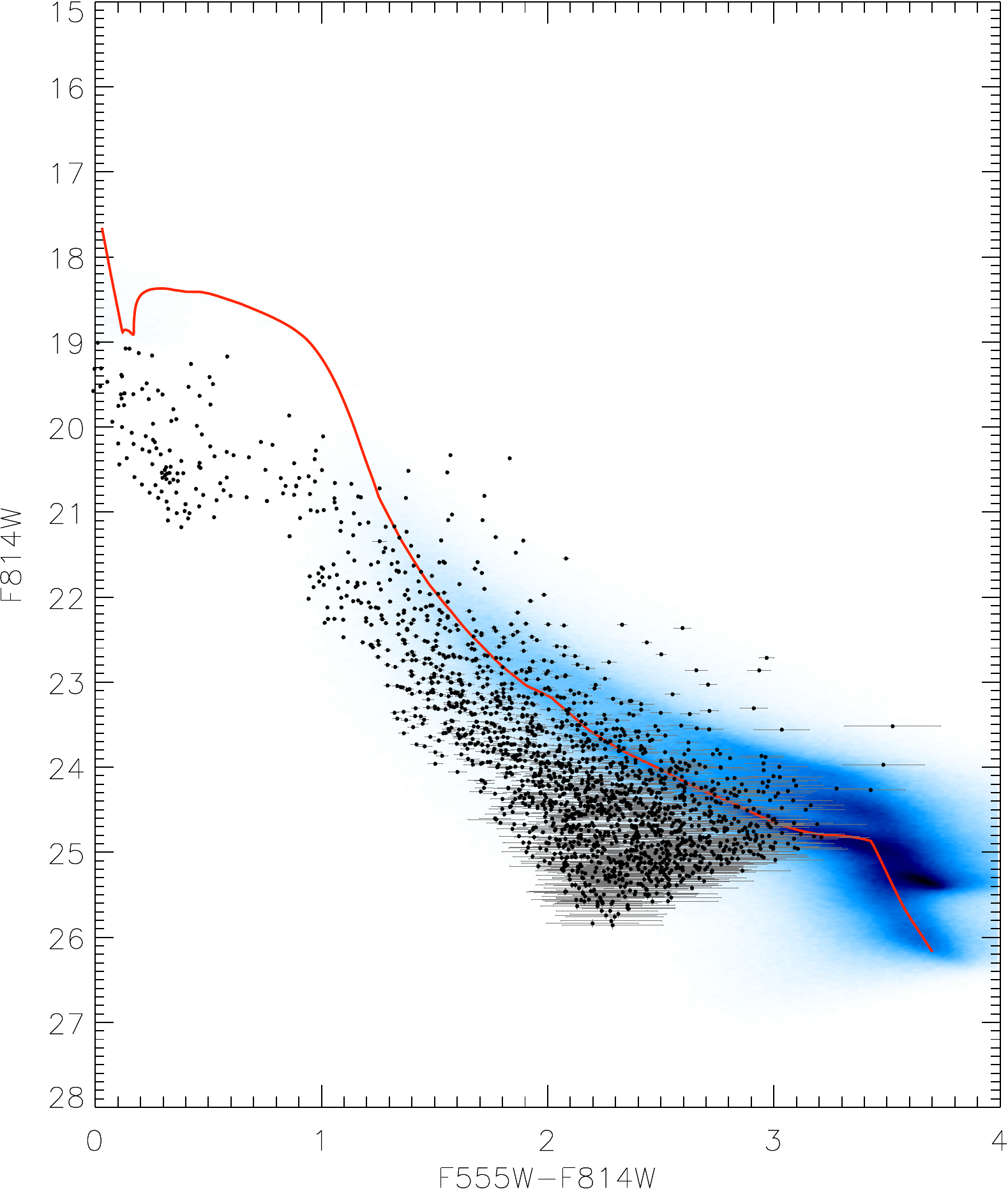}
\caption{Example of a 2~Myr isochrone from the {\sl Siess} models, which is converted into a 2D density CMD distribution (blue contour),
after the application of binarity (for an assumed binary fraction $f=0.5$), accretion, variability, differential extinction and crowding, as described in
\S~\ref{section:2Disochrones}. The original isochrone is overlaid with the red line. The dots are the stars of our photometry with error
bars indicating the $1\sigma$ photometric errors. An average reddening of $E(V-I) = 0.275$~mag, typical for LH~95 (Paper~I),
is applied. \label{fig:2Disochrone_example}}
\end{figure}

In Fig.~\ref{fig:2Disochrone_example} we show an example of the result of the application of our simulations to a 2~Myr isochrone
from the {\sl Siess} family of models. In this figure it is evident that the CMD-spread of a `perfect' coeval PMS stellar population, produced by the
biasses discussed above, can generate an observed dislocation of the stars of more than 1~mag in luminosity and up to 1~mag in color.
The bifurcation seen at the faint end of the simulated distribution is an effect of binarity. This is due to the fact that the evolutionary
models do not cover masses lower than $0.1$~M$_\odot$. As a consequence our random selection of binary pairs is unable to simulate
binaries with the second component below this limit. This effect does not alter the results of our analysis, because, as it is seen in
Fig.~\ref{fig:2Disochrone_example}, our photometry does not reach these masses.

\subsection{Derivation of stellar ages. The single star case.}
\label{section:deriving_single_ages}

In this section we use statistics to address the issue of deriving {\sl the most likely age of a single PMS star, for which the only
information is its measured brightness and the corresponding photometric errors}.
Specifically, we derive the age of each star in our catalog using a maximum-likelihood method by comparing the observed position of each star in the
CMD with a complete set of simulated 2D stellar distributions, constructed from theoretical isochrones with the
method described above. Each of these simulated distributions provides an estimate of the probability of finding a star
in a specific CMD location, providing that we know its age. In Section \ref{section:deriving_cluster_age} we will then generalize this approach to derive the best age of the entire cluster, considering all the measured stars simultaneously.

In practice, if the photometric errors were zero, the most likely age would be that of the modeled distribution
that gives the highest probability in the CMD position of the star. In reality, since there are errors in the measured
$m_{\rm 555}$ and $m_{\rm 814}$ magnitudes of the star, the observed position of the star in the CMD is also probabilistic, defined by
the gaussian distributions of the photometric errors. Therefore, if we consider a star observed in the position
$(V_i,I_i)$ on the CMD and a model stellar distribution of a given age $t$, for every point $(V_j,I_k)$ of the
CMD-space there are two probabilities that should be specified: a) the probability, defined by the 2D model,
that a star of age $t$ is located in  $(V_j,I_k)$, and b) the probability, defined by the Gaussian distributions
of the photometric errors, that the star we observe in $(V_i,I_i)$ is instead located in $(V_j,I_k)$ but displaced
by the photometric errors. The product of these two probabilities, integrated over the entire CMD-space, provides the likelihood that the considered star has an age equal to $t$. This likelihood, normalized on all the possible ages, represents the exact probability that the star has an age between $t$ and $t+{\rm d}t$.

\begin{figure}[t!]
\includegraphics[width=\columnwidth]{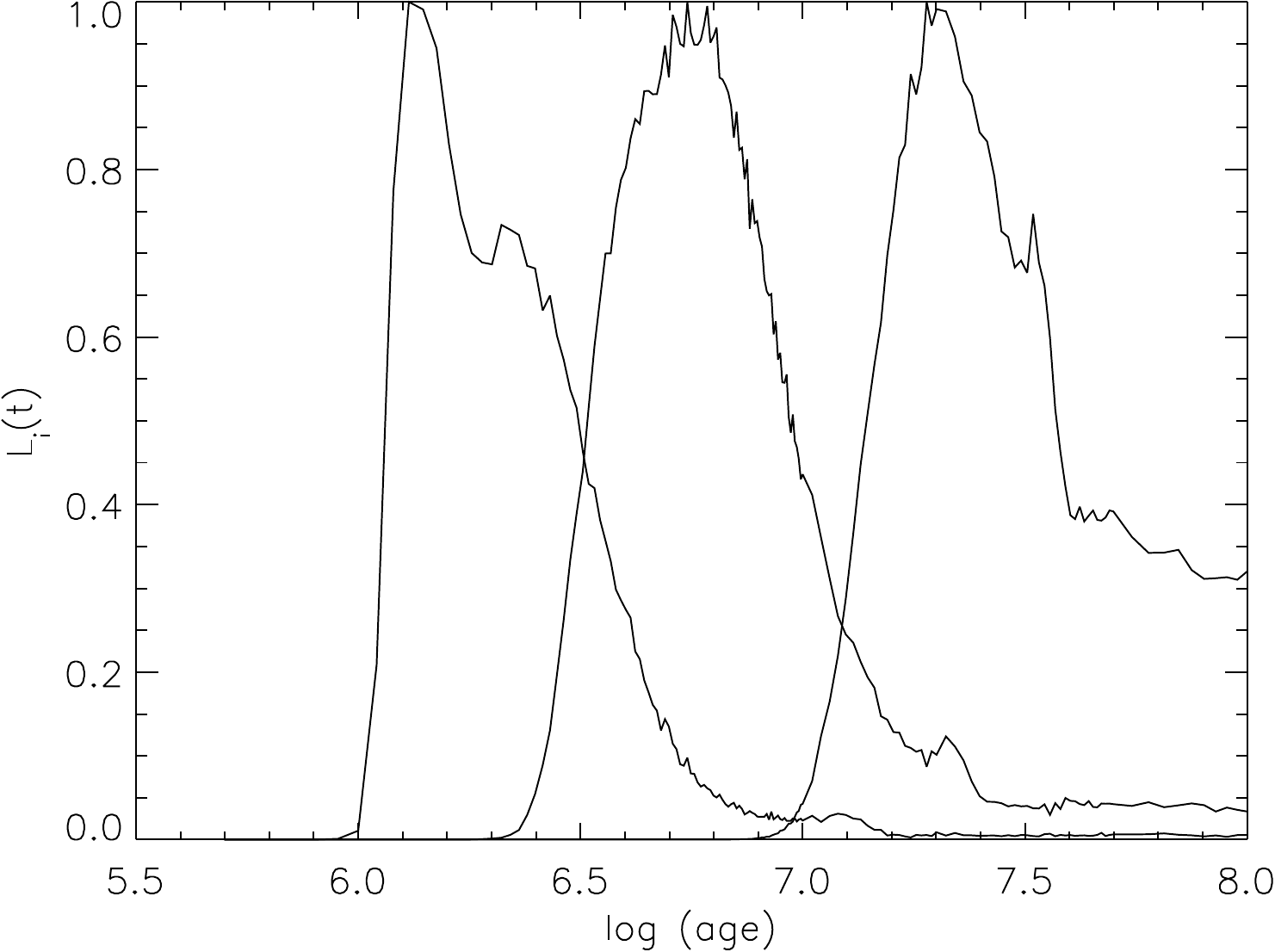}
\caption{Example of likelihood functions, $L_{i}(t)$, for 3 stars from our photometric
sample. The maxima of the functions indicate the assigned best-fit ages for the
sources, equal to 1.3, 5.5 and 19~Myr respectively in this example.
\label{fig:Li_function_example}}
\end{figure}

\begin{figure*}[t!]
\includegraphics[width=\columnwidth]{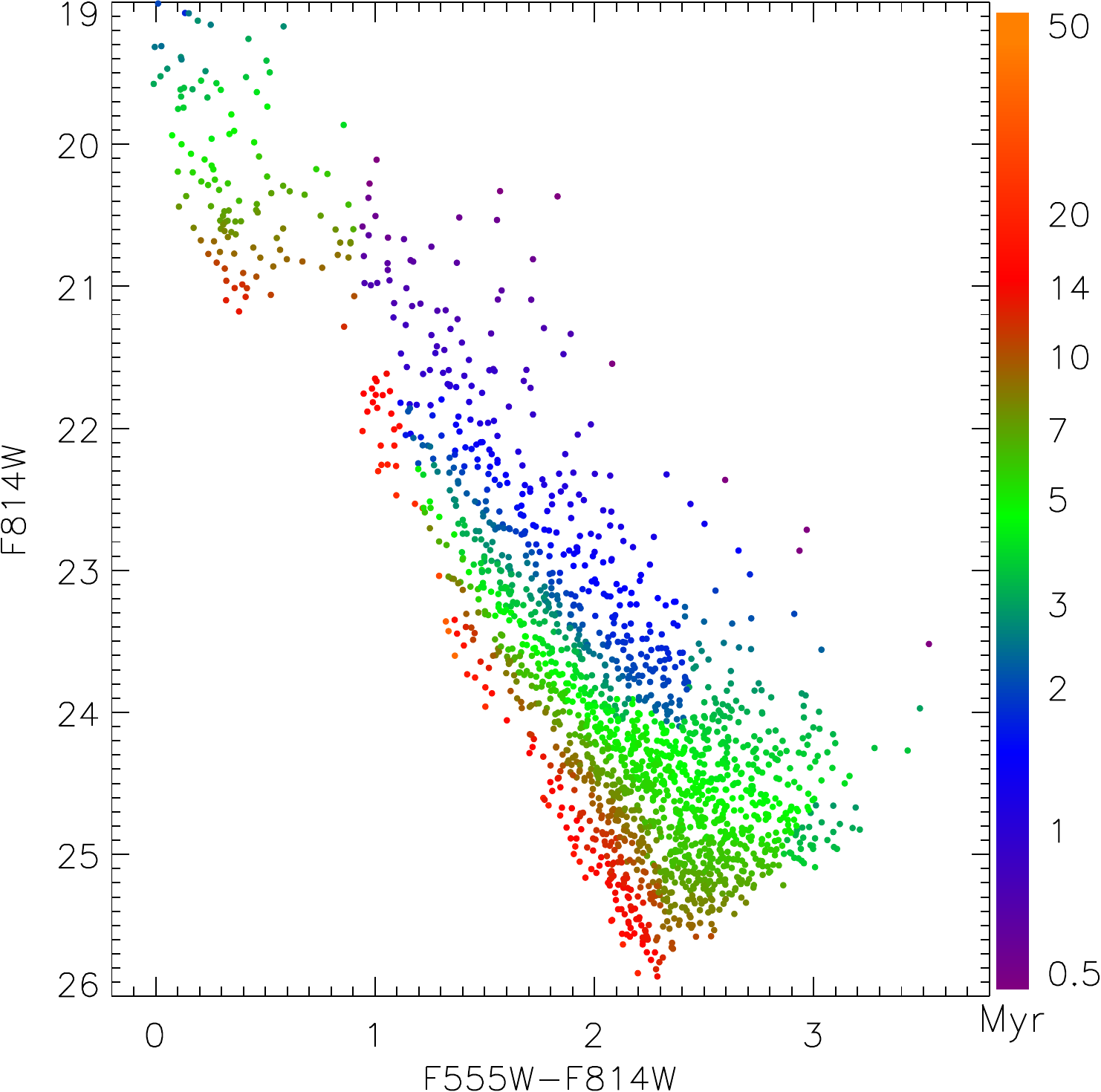}
\includegraphics[width=\columnwidth]{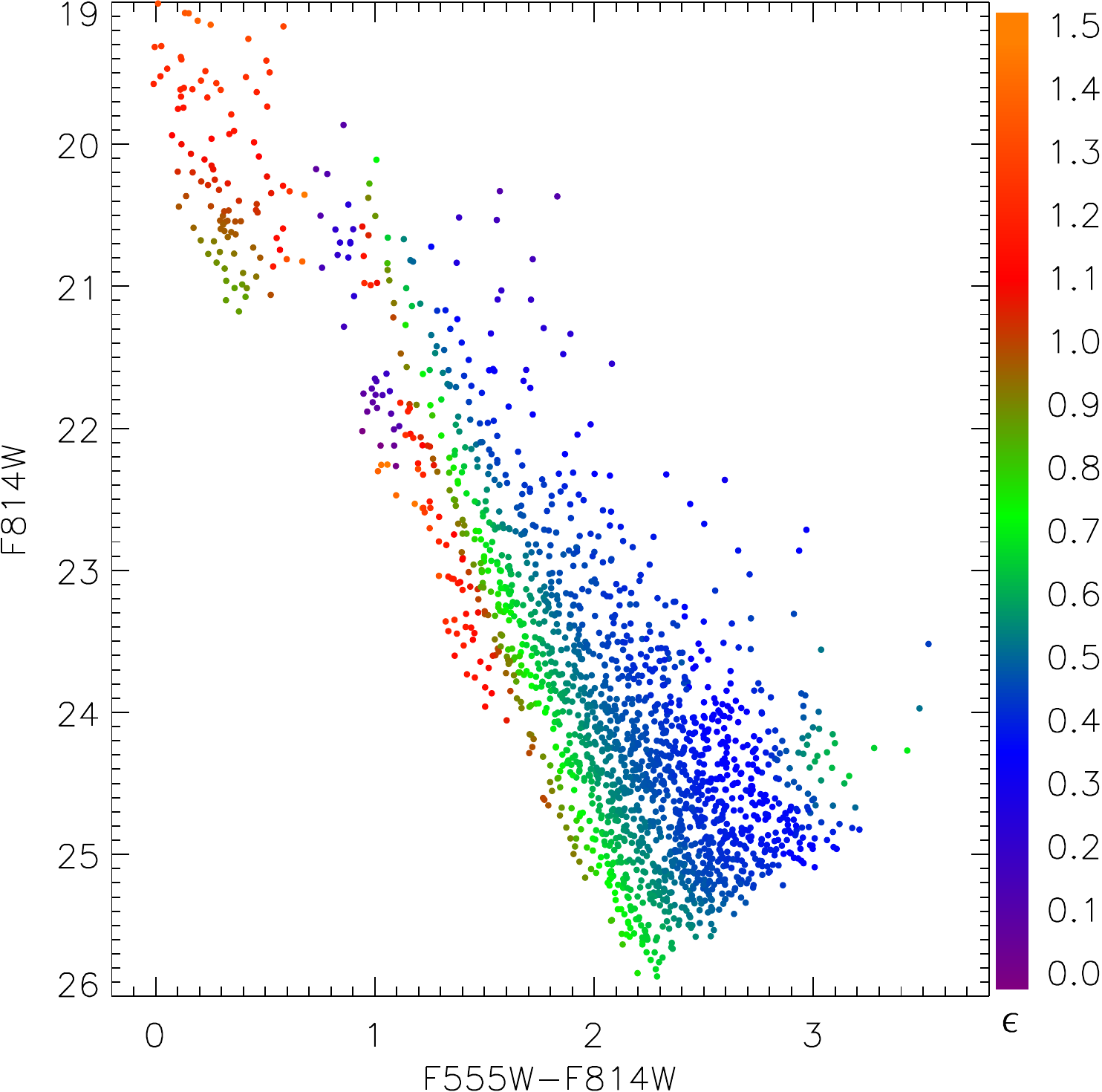}
\caption{\emph{Left}: Color-magnitude diagram with the PMS stars color-coded according to their individual
ages derived with our maximum-likelihood method. The model distributions were derived with the use of theoretical
isochrones from the {\sc FRANEC} family of models. The assumed binary fraction is $f=0.5$. \emph{Right}: The same
CMD with the stars color-coded according to their $\epsilon$ parameter (see \S~\ref{section:deriving_single_ages}). \label{fig:CMD_mostlikelyages}}
\end{figure*}

Since the photometric errors in the color-magnitude space are correlated, we apply this method in the magnitude-magnitude plane.
Mathematically, we define the likelihood $L_{i}(t)$ for the $i-$th star in our sample as function of the age $t$ as:
\begin{equation}
L_i(t) = \int_{V,I}\rho(V,I,t)\ U_i(V-V_i,I-I_i) {\rm d}V{\rm d}I
\label{equation:Li_t}
\end{equation}
\noindent
where $\rho(V,I,t)$ is the surface density computed from the isochrone of age $t$, and
\begin{equation}
\label{equation:gaussian_errors}
U_i(V-V_i,I-I_i)=\frac{1}{2\pi\sigma_{V_{i}}\sigma_{I_{i}}}e^{-\big[\big(\frac{V-V_i}{\sigma_{V_{i}}}\big)^2+\big(\frac{I-I_i}{\sigma_{I_{i}}}\big)^2\big]}
\end{equation}
\noindent
is the 2D Gaussian representing the photometric error of the $i-$th star.

In Fig.~\ref{fig:Li_function_example} we show an example of the likelihood function $L_i(t)$ computed numerically
for three stars in the LH~95 catalog, assuming 2D models obtained from the {\sc FRANEC} grid and assuming a binary fraction $f=0.5$.
The age assignment for every star is performed by maximizing $L_i(t)$. For example, the three stars whose $L_i(t)$ functions are shown in Figure \ref{fig:Li_function_example}, will have most probable ages of 1.3~Myr, 5.5~Myr and 19~Myr.

In order to estimate the accuracy in the age determinations, we isolate for every likelihood function the range, in $\log{(t)}$
that includes 68\% of the area under the curve, in analogy with the $\pm1\sigma$ interval for a normal distribution. Since for
some stars the shape of  $L_i(t)$ is highly skewed, we choose to define the limits of the interval to delimit identical areas of
34\% of the total to the left and to the right from the peak of the distribution. We then call $\epsilon$ the width of the range, in units
of $\log{(t)}$. Indicatively, if a $L_i(t)$ were normal, $\sigma(t)$ would be equal to $\epsilon/2$.

Fig.~\ref{fig:CMD_mostlikelyages}({\em left}) shows the derived best-fit stellar ages in the CMD, assuming the evolutionary models
of {\sc FRANEC} and a binary fraction $f=0.5$. In Fig.~\ref{fig:CMD_mostlikelyages}({\em right}) we show the corresponding variation
of $\epsilon$ along the CMD. It is evident that, whereas the age assignment turns out to be relatively precise in the
PMS regime, for higher masses, close to the PMS-MS transition ($m_{\rm 814}\lesssim21$) the uncertainty increases up to
$\epsilon\simeq1.5$~dex. This is  due to the fact that PMS isochrones of different ages are closer to each other in this range,
and tend to overlap in the MS regime. The higher values of $\epsilon$ on the blue side of the sequence are also due to the
fact that theoretical isochrones of ages $\gtrsim10$ are also close to each other, since PMS evolution is significantly slower for these ages than for younger stages of PMS contraction.

\begin{figure}
\includegraphics[width=\columnwidth]{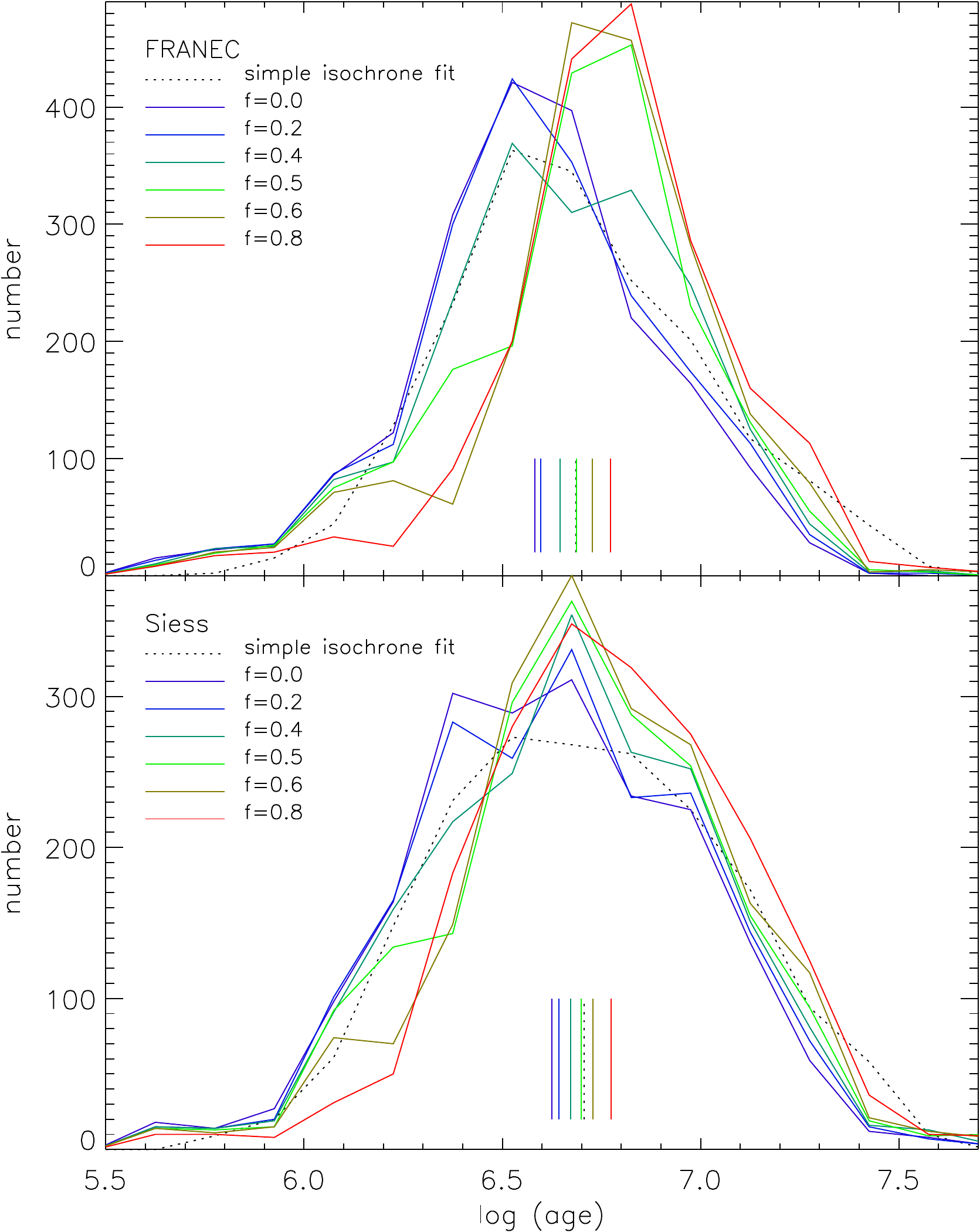}
\caption{Distribution of maximum-likelihood stellar ages for the LH~95 members, computed using
2D CMD distribution models, based on {\sc FRANEC} (upper panel) and {\sl Siess} (lower panel)
evolutionary models. We assume different binary fraction values $f$ from 0.0 to 0.8. For each
family of isochrones, we also show the distribution of ages computed only by interpolation of
isochrones in the CMD, without accounting for physical and observational biases (black lines) .
The vertical lines (colored appropriately) in each panel indicate the mean value of the age for
each plotted distribution. \label{fig:histogram_singlestars_bestages}}
\end{figure}

We compare the distributions of the best-fit ages for our photometric sample changing the binary fraction and the original family of
evolutionary models. This is shown in Fig.~\ref{fig:histogram_singlestars_bestages}, where we also indicate the average age for
each distribution. It this figure is evident that increasing the binary fraction leads to the prediction of older ages. This confirms the findings of \citet{naylor09}.
Indeed, a higher fraction of binary systems produces a simulated 2D distribution shifted towards higher luminosities, hence older measured ages. Passing from $f=0$ to $f=0.8$, we predict a difference in age of up to 0.2~dex (which is 60\% older ages.).

In the same figure we also overlay the distribution of ages computed only by interpolation between theoretical isochrones, without the addition of any observational and physical bias except for the average extinction of LH~95.
In this case we find that the average age is higher than that derived from our statistical method assuming $f=0$. This is due to the effect of dark spots and accretion. The first biases
the predicted luminosities towards fainter values; the second towards brighter luminosities and bluer colors. In particular, the color effect caused by veiling affects more the position in the CMD than the increase in the fluxes, requiring an older isochrone to fit the observed sequence.

In conclusion, we find that, in the derivation of stellar ages for a young PMS cluster, neglecting binaries leads to an underestimation of the ages, neglecting variability and accretion to an overestimation of those. The two effect counterbalance, in our case, for a binary fraction $f=0.5$ and assuming star-spot and accretion values from the 2~Myr Orion Nebula Cluster.

\subsection{Spatial variability of stellar ages}
\label{section:spatial_variability}
\begin{figure}
\includegraphics[width=\columnwidth]{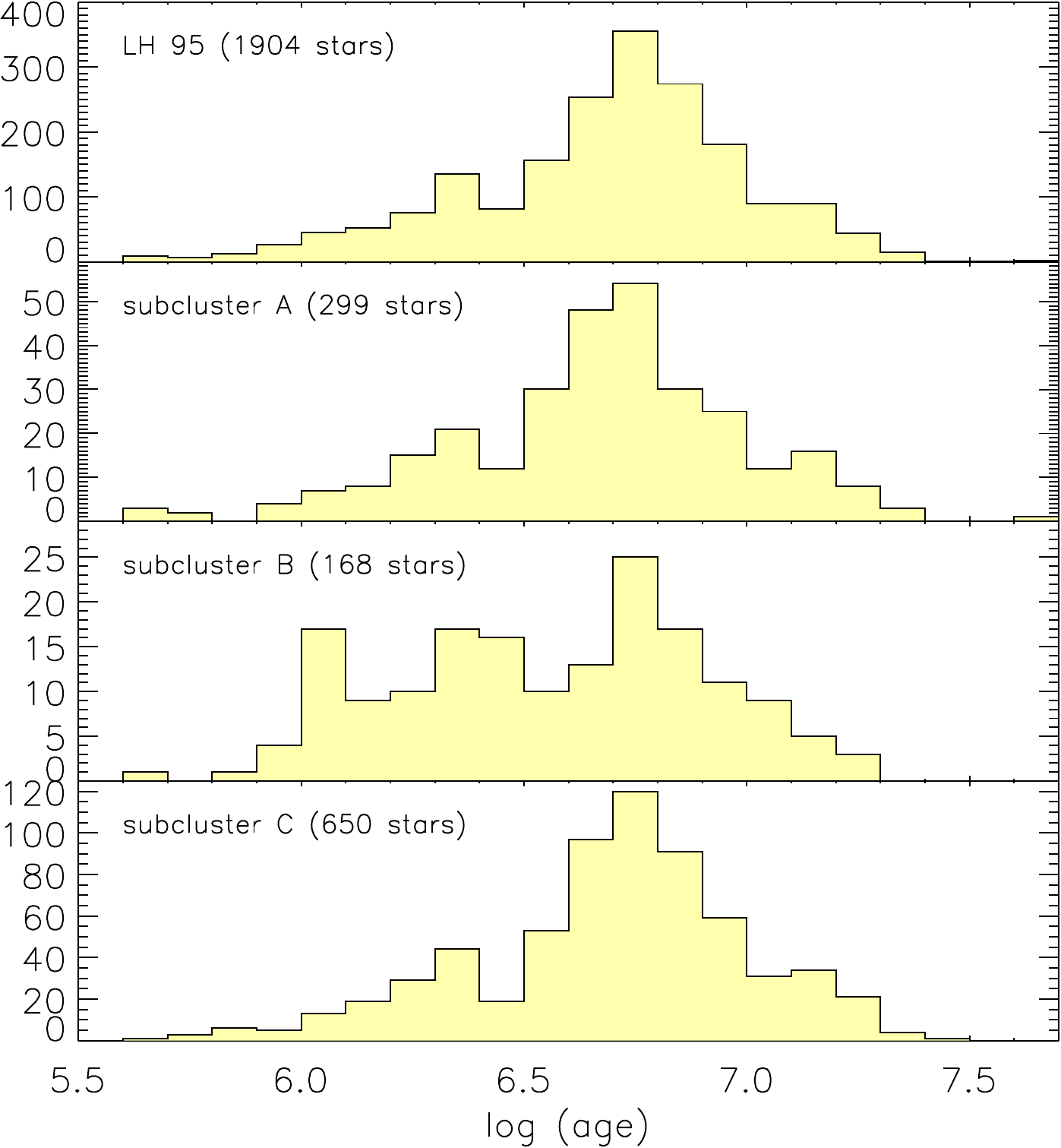}
\caption{Comparison between the age distribution of the entire region and the three denser sub-clusters, for {\sc FRANEC} evolutionary models and $f=0.5$.
 \label{fig:histogram_singleages_subsclusters}}
\end{figure}
As we mentioned in Section \ref{s:dataset}, LH~95 is not a centrally concentrated cluster. It presents 3 main concentrations of PMS stars, named in our previous work ``subcluster A, B, C'', located respectively at the eastern, western, and southern edges of LH95.  They include about half of the stellar population of the region, while the rest of the population is distributed over the area between the three. Since the projected distance between the stellar clumps is quite large (15 - 20~pc), it could be plausible that star formation did not occur simultaneously all over the entire region, but sequentially. A Galactic example of such scenario can be found in the Orion complex \citep{briceno2008}, where different populations, from 2 to 10~Myr old, are distributed over a similar range of distances.

We investigate if the stellar sub-groups in LH95 present different average ages, in order to rule out the possibility that the measured broad age distribution for the entire region (Figure \ref{fig:histogram_singlestars_bestages}) is simply due to a superposition of populations with different ages, instead of the effect of observational biases and, maybe, a real age spread.
To this purpose we isolate the members in the three subclusters, using the same selections as in Paper~I. Then we consider the distribution of ages limited to each of the three subregions, in comparison with that of the entire LH~95. This is illustrated in Figure \ref{fig:histogram_singleages_subsclusters}. It is evident that, in each region, the distributions are very similar, spanning from less than 1~Myr to 25~Myr, with a peak at $\sim$5~Myr. The only evident anomaly is an overabundance of young members for sub-cluster B. However, this group of stars encloses the densest concentration of PMS stars in the whole region, with more than 150 star within $\sim2$~pc. Because of this, stellar confusion due to crowding should be particularly dominant in this area, producing a number of overluminous stars due to unresolved superposition, and, thus, a fraction of apparently younger members. This is, however, not a problem for the analysis of the cluster age (Section \ref{section:deriving_cluster_age}). In our CMD modeling (Section \ref{section:2Disochrones}) we have purposefully included the effect of crowding affecting 5\% of members. The overabundance of young stars in subcluster B is caused by about 40 stars, which are $\sim2\%$ of the total stellar sample.

Considering this, we exclude the presence of any spatial variations of the ages in LH~95. Despite the irregular density of stars and the relatively large size of the region, LH95 must have formed as a whole, in a unique episode of star formation.

\subsection{Age of the entire cluster}
\label{section:deriving_cluster_age}

In section \ref{section:deriving_single_ages} we derived the most likely age for every star in the sample.
Here we derive the best-fit age for the entire stellar system. In a similar way as in NJ06, we compute this age by
maximizing the ``global'' likelihood function $L(t)=\prod_iL_i(t)$. To compute $L(t)$ we restrict our treatment to the stellar
sample with $22\leq m_{\rm 555}\leq 27$. In particular, we want to eliminate any biases introduced by detection incompleteness, which
would lead to the detection at the faint-end of more young bright members than old faint ones. The detection completeness
reaches $\sim90\%$ at $m_{\rm 555}=27$ (Paper I), and we consider this threshold as a reasonable limit for our analysis here.
We stress that, as in \citet{naylor09}, we could include the detection completeness function in the 2D models derived in Section \ref{section:2Disochrones}, and then  recover the implied cluster properties using our maximum likelihood procedure considering all the stars. In this way, we could test for bias also as a function of our chosen completeness limit. Nevertheless, considering the
large stellar sample we have at disposal, we prefer to limit out analysis on the luminosity range where we that detection incompleteness does not affect our photometry.
In this way we remove any additional uncertainty in the cluster age caused by a possible inaccuracy of the measured completeness function.

In addition, we do not include in the analysis the bright members of  both the MS and the PMS-MS transition, for two reasons: 1) as shown
in Fig.~\ref{fig:CMD_mostlikelyages}({\sl right}), the accuracy of the age estimation for these sources is poor, and 2) these sources
have very small photometric errors, which means that the 2D Gaussian distribution representing the errors
(Eq.~\ref{equation:gaussian_errors}) goes to zero very fast in the neighborhood of the specific CMD loci, and so does $L_i(t)$.
Such very small values of $L_i(t)$ would dominate the product $\prod L_i(t)$, biasing the resulting global likelihood function.
From a mathematical point of view this effect is legitimate, because it only means that during the fit the data points with the
smallest uncertainties dominate the statistics. However, in practice if the observational errors are much smaller than the intrinsic
uncertainty of the models (in our case, the uncertainty in the theoretical evolutionary model computation and the modeling of
the scattering) a strict application of the former can lead to unrealistic results.

\begin{figure}[t!]
\includegraphics[width=\columnwidth]{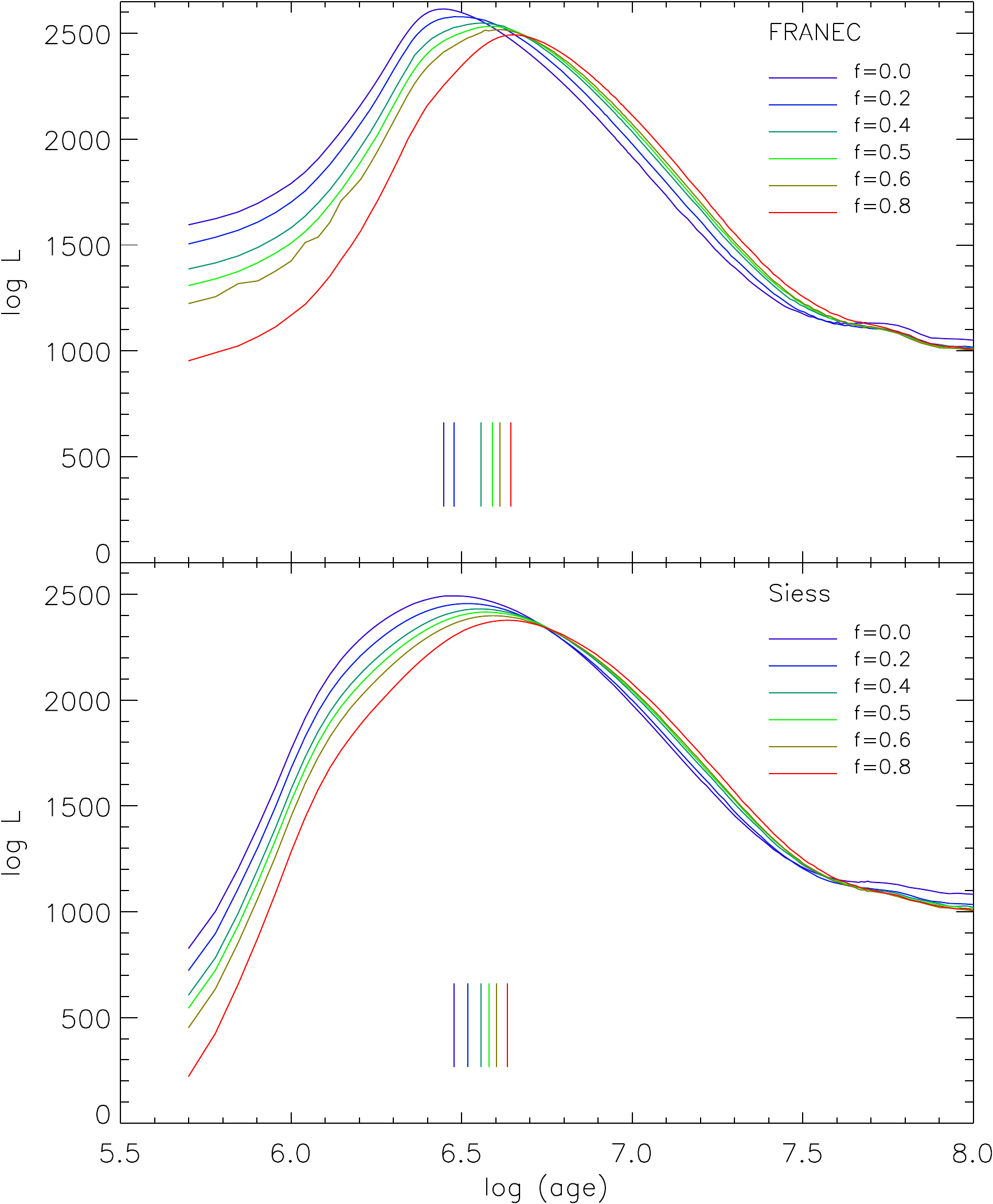}
\caption{Global likelihood functions $L(t)$, derived from the simultaneous fit of the the 2D stellar distribution models
to all PMS stars in our photometric sample with $22<m_{\rm 555}<27$ based on {\sc FRANEC} (upper panel) and {\sl Siess}
(lower panel) evolutionary models. We assume different binary fraction values $f$ from 0.0 to 0.8.
The vertical lines (colored appropriately) in each panel indicate the age of the maxima of each curve.
\label{fig:plot_globallikelihood}}
\end{figure}

In Fig.~\ref{fig:plot_globallikelihood} we show the derived $L(t)$ functions, in logarithmic units, for both {\sc FRANEC} and {\sl Siess}
isochrones and different choices of binary fraction $f$. The plateau at old ages, in comparison to the cut-off at young ages, is due
to the fact that old isochrones tend to be closer to each other and closer to the sequence of PMS stars observed in the CMD than
the young isochrones. As a consequence, the `distance' from a given star, in units of $\sigma$ of the photometric errors, is on
average lower for the older isochrones than for the younger, resulting to higher probabilities for the former.

Results of our analysis are given in Table~\ref{table:globalages}, where we show the ages at the peak of each $L(t)$, as well as the
associated values of likelihood for each choice of parameters.
Our analysis also shows that the best-fit age is very sensitive to the assumed binary fraction, consistently with what we found in Section \ref{section:deriving_single_ages} analyzing the distribution of ages for single stars.
 as it increases of up to 60\% with
increasing binarity (depending on the considered isochrone models). The peak values of $\log{L(t)}$ given in
Table~\ref{table:globalages} do not have an absolute meaning, given the arbitrariness of the normalization of the 2D density models
$\rho(V,I,t)$.

\begin{figure*}[t!]
\includegraphics[width=\columnwidth]{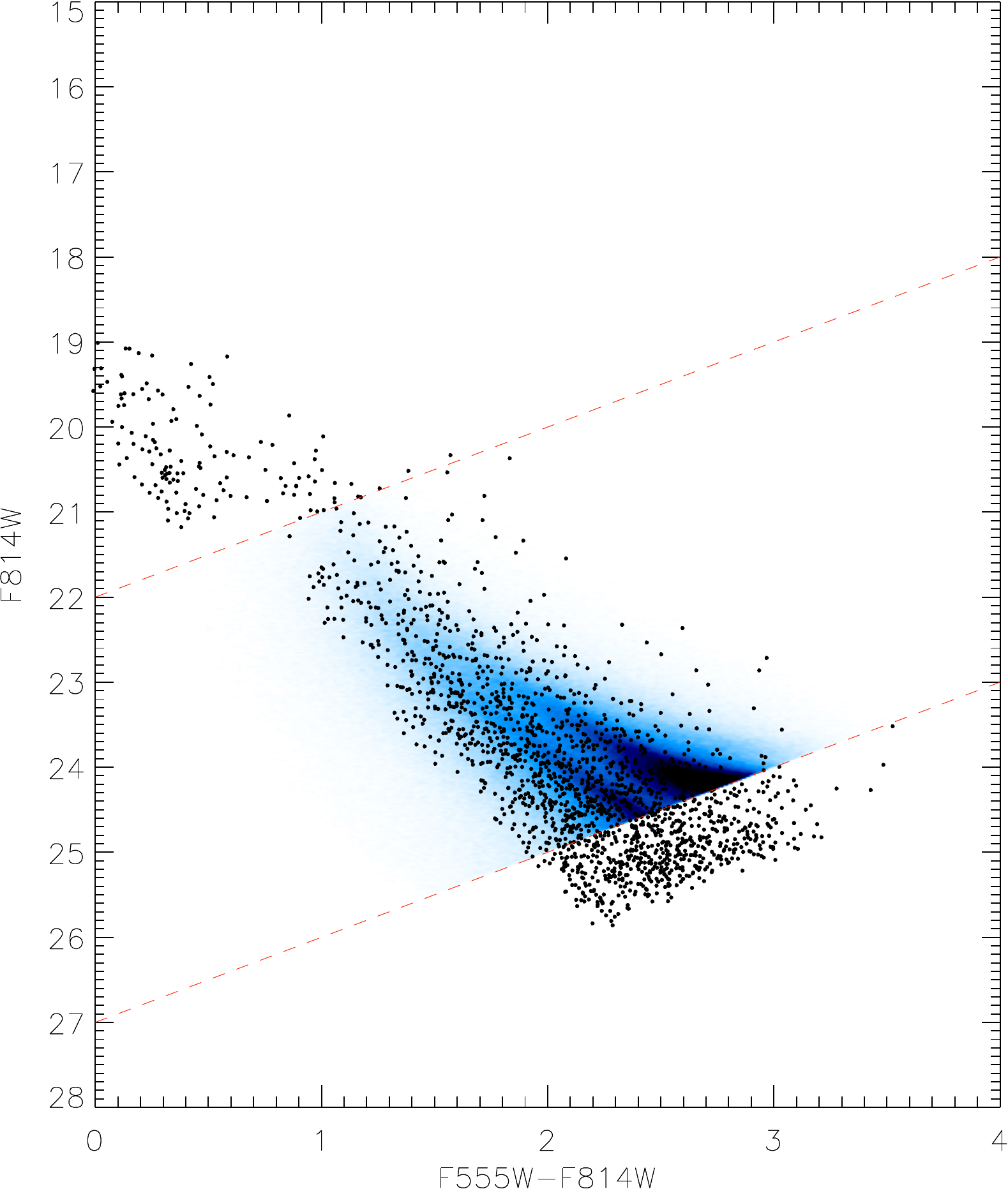}
\includegraphics[width=\columnwidth]{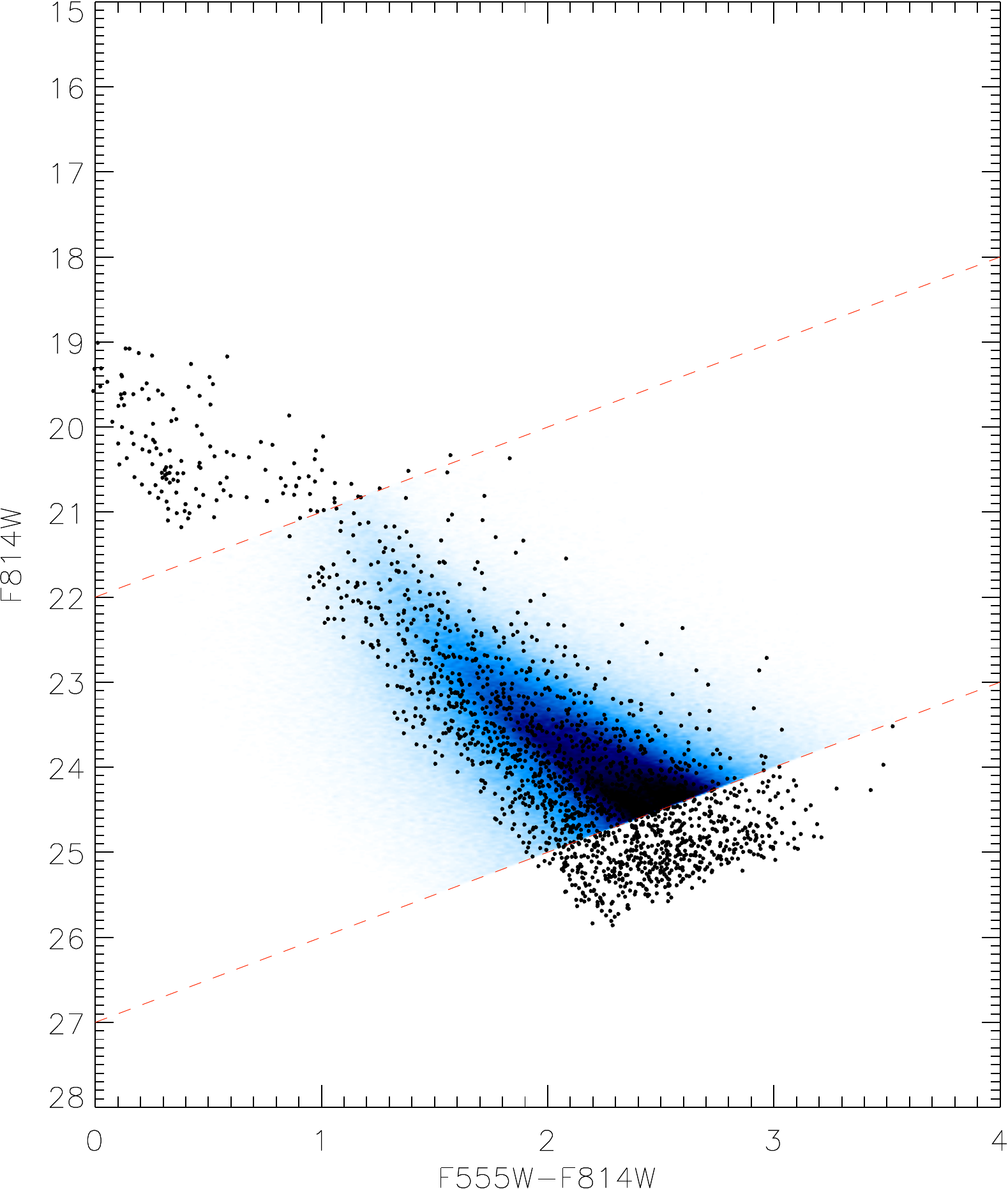}
\caption{The observed CMD with the 2D density models that provide the best fit to the observed sequence of PMS stars,
assuming binarity of $f=0.5$ for the {\sc FRANEC} ({\em left panel}, age 3.9~Myr) and  the {\sl Siess} ({\em right panel}, age
3.8~Myr) family of evolutionary models. The dashed lines denote the subsample of the photometry used in this analysis. For clarity, only the portion of the 2D models within this luminosity range is shown.
\label{fig:2Disochrone_example_best}}
\end{figure*}

In Fig.~\ref{fig:2Disochrone_example_best} we show again the observed CMD, with the best-fit 2D density models from {\sc FRANEC}
({\em left}) and {\sl Siess} ({\em right}) grids of isochrones superimposed for $f=0.5$, for the luminosity range we considered in our analysis.
The two models are very similar, with {\sc FRANEC} producing a slightly over-density at the reddest
end of the considered magnitude range.

\begin{deluxetable}{ccccc}
\tablecaption{Most-likely ages for the entire cluster.}
\tablehead{
\colhead{} &
\multicolumn{2}{c}{{\sc FRANEC} Models}  &
\multicolumn{2}{c}{{\sl Siess}  Models} \\
\colhead{$f$} &
\colhead{age (Myr)} &
\colhead{$\log{L_{\rm max}}$} &
\colhead{age (Myr)} &
\colhead{$\log{L_{\rm max}}$}
}
\startdata
   0.0 &     2.8 &  2615.26 &      3.0 &  2492.16 \\
   0.2 &     3.0 &  2578.42 &      3.3 &  2456.14 \\
   0.4 &     3.6 &  2548.63 &      3.6 &  2431.42 \\
   0.5 &     3.9 &  2533.37 &      3.8 &  2415.62 \\
   0.6 &     4.1 &  2517.49 &      4.0 &  2398.78 \\
   0.8 &     4.4 &  2492.71 &      4.3 &  2377.05
\enddata
\label{table:globalages}
\end{deluxetable}

From Figure \ref{fig:2Disochrone_example_best}, it is evident that an additional spread, besides that modeled in the 2D density distribution, is evident in our data. In particular, there are observed members at the sides of the sequence which may not be well modeled by the simulated distributions for a single cluster age. This may be indicating of an additional age spread in LH95. In the next sections we study this possibility.

\section{Age-spread of PMS stars in LH~95}
\label{section:deriving_age_spread}

\subsection{Verification of an age-spread}

In the previous section we have determined the average age of LH~95 isolating the 2D model that provides the best fit to the observed
sequence, for two families of theoretical evolutionary models. In this section we wish to investigate whether the spread of the PMS stars
in the observed CMD is wider or not than that of the simulated 2D density distributions as they are produced by accounting for the sources
of CMD-broadening discussed in \S~\ref{section:2Disochrones}, {\sl after we additionally apply the photometric uncertainties}. If indeed the
observed CMD broadening of PMS stars is wider than the synthetic derived from a single isochrone, under the hypothesis that we do not have underestimated the sources of apparent luminosity spread in the CMD, this would suggests the presence of
a {\em real} age-spread in LH~95. The method applied in this section, described below, also allows us to qualify the goodness of our fit.

This method can be compared to a standard $\chi^2$ minimization, in which the best fit model is obtained by minimizing
the $\chi^2$, and this value is then compared to the expected distribution of $\chi^2$ given the degrees of freedom. A measured
$\chi^2$ which significantly exceeds the expected distribution indicates a poor fit, while a  value too low suggests that the fit is
suspiciously good, which is the case, e.g., of overestimated errors. In our statistical framework, the procedure is qualitatively
similar, with two main differences: i) since the measured $\log L(t)$ is maximized, a poor fit is found when the measured value
is lower than expected; ii) the expected distribution of $\log L(t_{\rm best-fit})$ is not analytical but can be derived numerically.
In general, this distribution varies for different assumed 2D models, and depends on the number of data points.

We compute the expected distribution of $\log L$ as follows. We consider the two best-fit 2D models shown in
Fig.~\ref{fig:2Disochrone_example_best}, limiting ourselves, for the moment, to the case of an assumed binary fraction $f=0.5$. We limit
the model to the same range used to derive the cluster age ($22\leq m_{\rm 555} \leq 27$), and simulate a population of
stars randomly drawn from the distribution of the model itself, as numerous as the LH~95 members in the same luminosity
range. We consider each of the simulated stars, and displace its position in the CMD by applying the average
photometric error measured in its immediate CMD area. In particular, for each simulated source we consider the 60 observed stars closer to it in the magnitude-magnitude plane. Than we randomly assign the photometric error of one of these 60, with a weight proportional to the inverse of the distance in magnitudes.
We then apply the same method described in
\S~\ref{section:deriving_cluster_age} to the simulated population, deriving $\log L$. We iterate this procedure 100 times,
each time randomly simulating a different population of test-stars from the same best-fit 2D model. In this way we obtain 100
values of $\log L$ describing the \emph{expected distribution} of this variable for the considered model, which is compared to the
\emph{measured} value computed for the observed sequence of PMS stars.

For the two models shown in Fig.~\ref{fig:2Disochrone_example_best}, i.e., the {\sc FRANEC} 3.9~Myr and the {\sl Siess} 3.8~Myr
models for $f=0.5$, we obtain predicted $\log L$ values of $2739 \pm 20$ and $2473 \pm 19$ respectively ($\pm$1$\sigma$ interval). In
comparison, the measured values of 2533 and 2416 (see Table~\ref{table:globalages}) are significantly lower, at $\sim-10\sigma$ and $-3\sigma$ from the expected values. This implies that the
observed sequence {\sl cannot originate from the best-fit models by additionally applying  photometric errors}, but that the stars
tend to be statistically located in regions of the CMD where the 2D model has lower density $\rho$ (see Eq.~\ref{equation:Li_t}),
and hence a lower average $\log L_i$ for the stars. For guidance, the low density regions of the 2D models in
Fig.~\ref{fig:2Disochrone_example_best} are those indicated with lighter blue colors, i.e., at high luminosity, where the IMF predicts
fewer members, and at both sides along the 2D distribution in the CMD, at large distance from the peaks of the 2D sequence itself.
Since the IMF we use to compute the 2D models has been derived from the actual observed data, we exclude the possibility
that the low measured value of $\log L$ is due to an overabundance of intermediate mass stars in comparison to low-mass stars.
As a consequence, our results suggest that the observed sequence of PMS stars in the CMD {\sl is broader than that predicted
by the 2D model for a single age}. There are two possible interpretations for this: a) LH~95 is formed by a coeval population, and the extra spread in the observed CMD is due to additional sources of apparent broadening in the CMD besides those we have included in our models; b) LH~95 hosts a real age-spread among its PMS stars. From hereafter we will assume the second possibility, and, under the validity of our modeling, we determine the age spread of the population.

\begin{figure}
\includegraphics[width=\columnwidth]{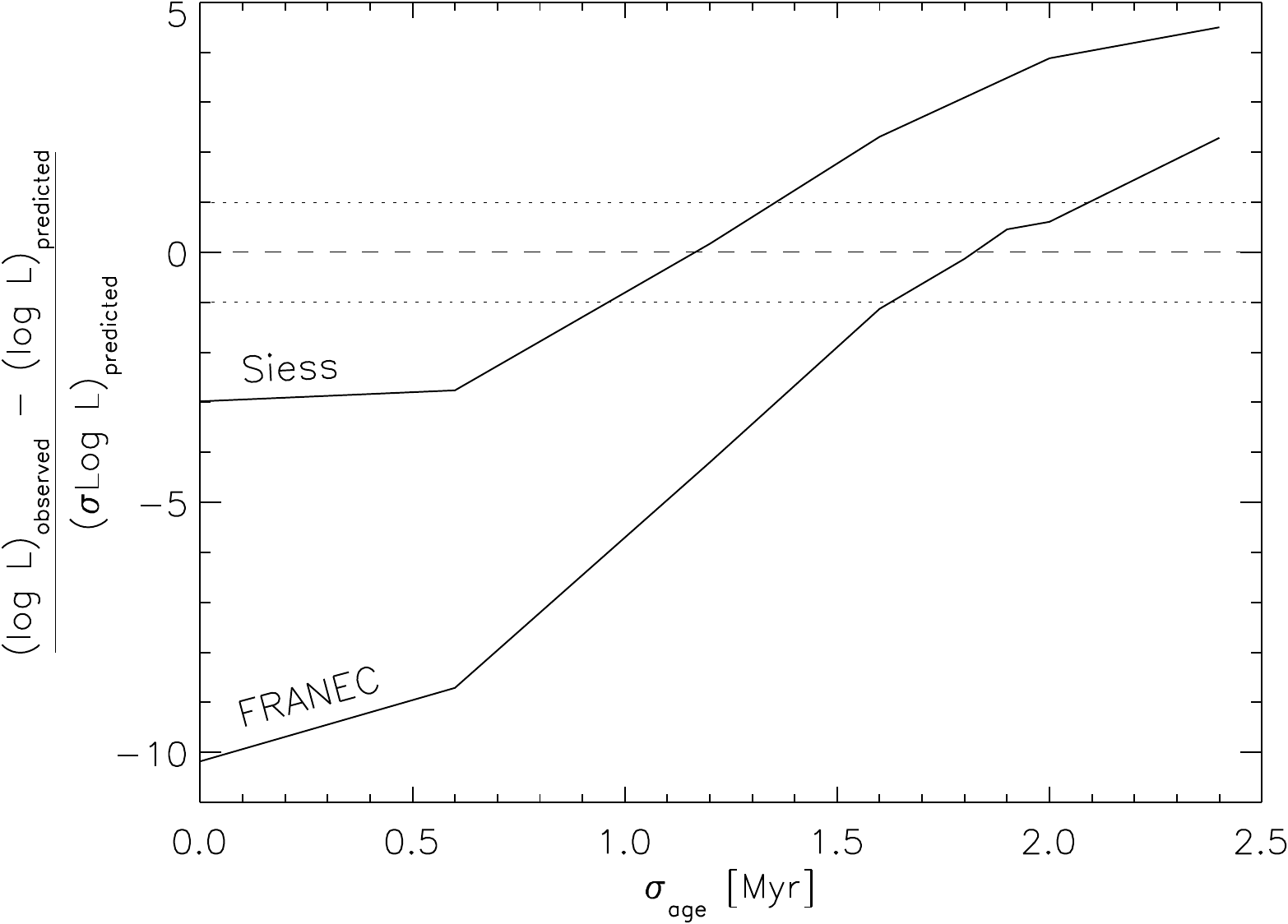}
\caption{Comparison between the measured value of log-likelihood and the predicted one, in units of standard deviation of the latter, as a function of the age spread.\label{fig:improvement_in_logl_with_spread}}
\end{figure}
\begin{figure*}
\includegraphics[width=\columnwidth]{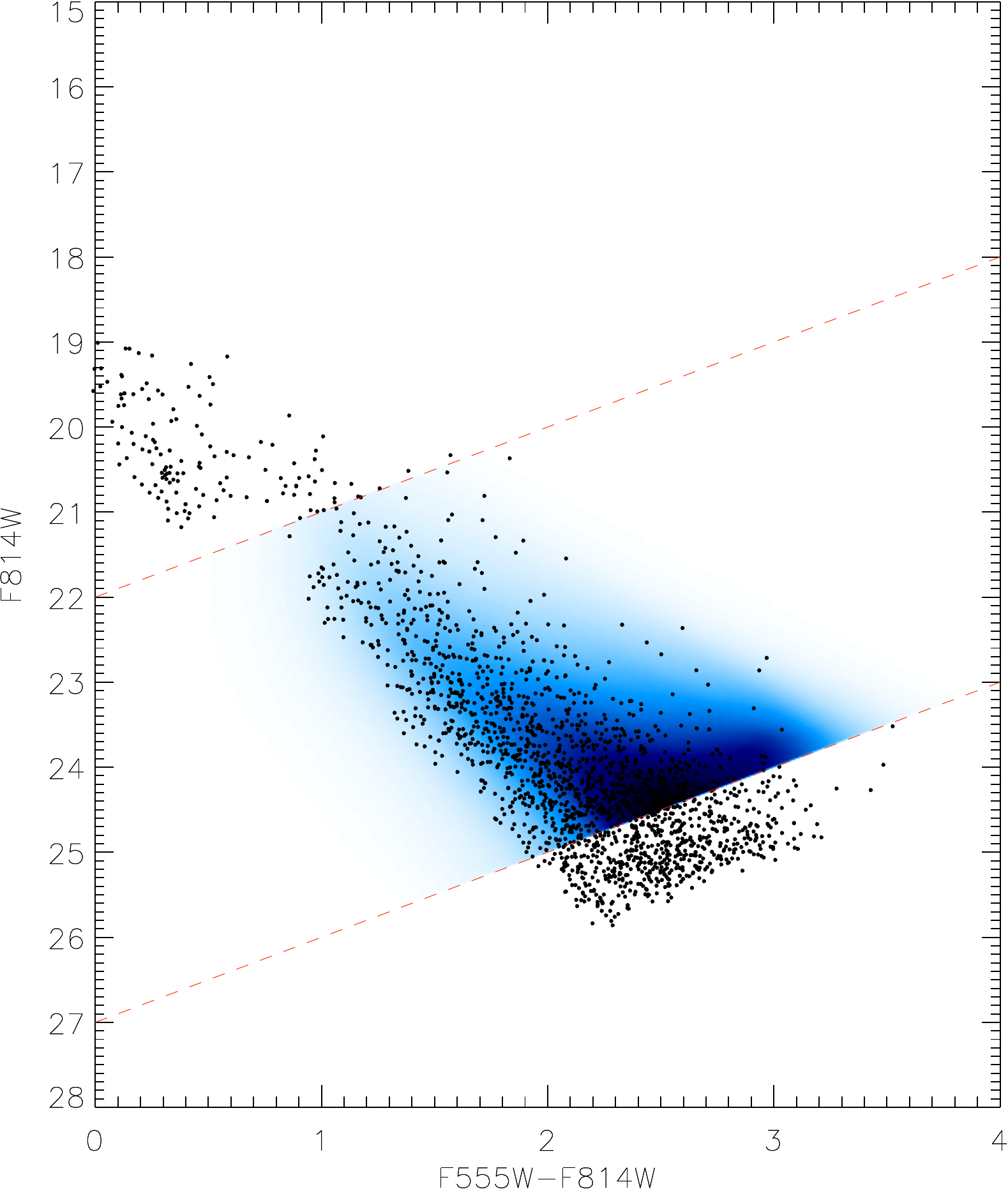}
\includegraphics[width=\columnwidth]{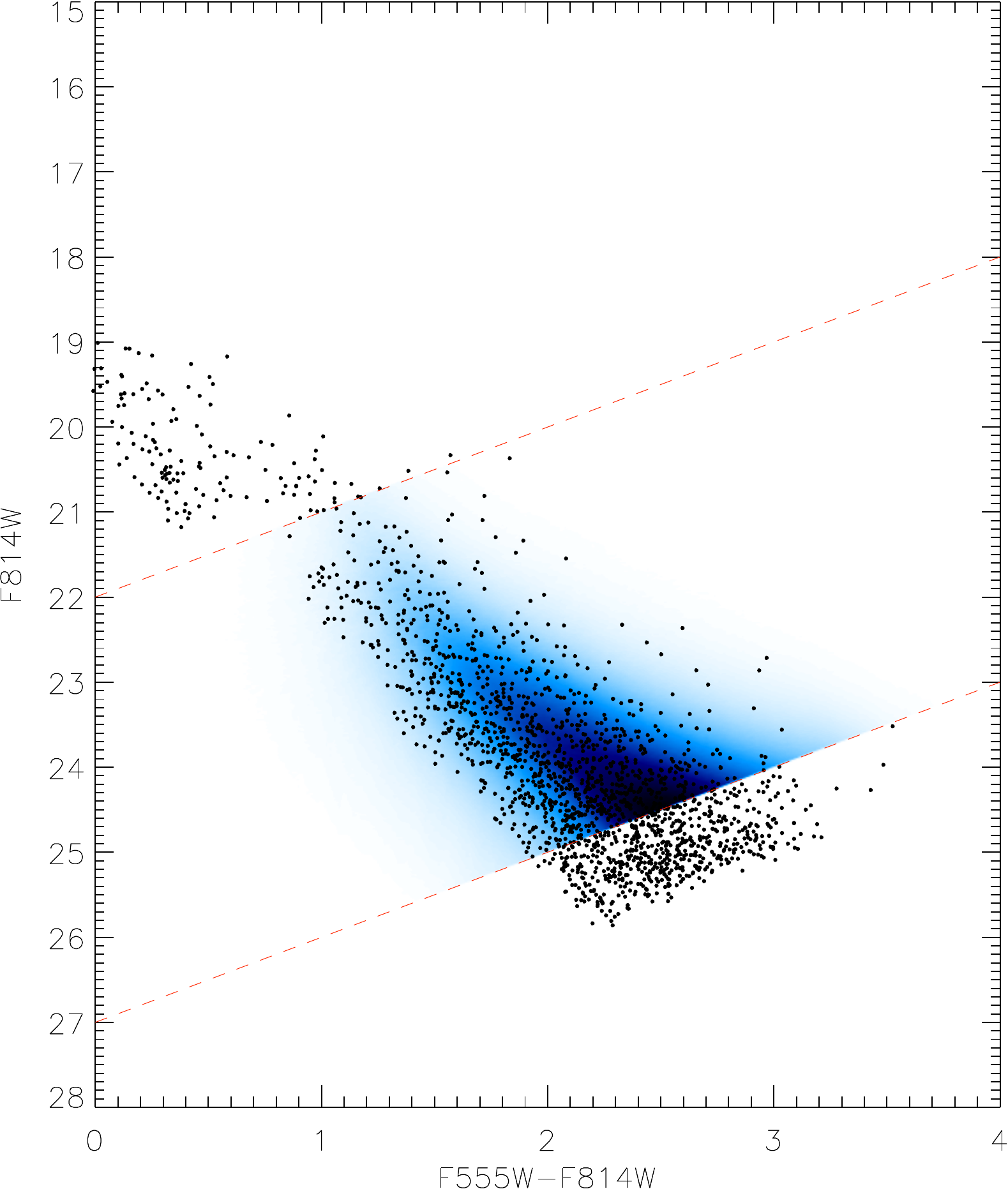}
\caption{The observed CMD with the 2D density models that provide the best fit to the observed sequence of PMS stars,
including an intrinsic age-spread, for an assumed binarity of $f=0.5$. \emph{Left}: The model obtained from {\sc FRANEC}
isochrones, with an average age
of 3.8~Myr and a Gaussian age-spread of $\sigma_{\rm age}=0.18$~Myr.  \emph{Right}: The model obtained from {\sl Siess}
isochrones, with an average age of 3.9~Myr and $\sigma_{\rm age}=0.12$~Myr.  The dashed lines denote again the subsample of the
photometry used in the analysis. \label{fig:2Disochrone_example_bestspread}}
\end{figure*}

\subsection{Evaluation of the age-spread}
\label{section:deriving_age_spread2}
In order to quantify the age-spread we include in our construction of the 2D model distributions, apart from the sources of CMD-broadening
of PMS stars previously considered (\S~\ref{section:2Disochrones}), also a distribution in ages. Specifically, we consider the best fit age
previously measured for the cluster (Table~\ref{table:globalages}) along with a Gaussian distribution of ages peaked at the best fit age with
standard deviations varying from $\sigma_{\rm age}=0.1$Myr to $\sigma_{\rm age}=3$~Myr. We construct, thus, broader 2D model distributions by essentially adding
together different single-age models, computed in \S~\ref{section:2Disochrones}, each weighted according to the Gaussian
distribution. For each value of $\sigma_{\rm age}$ we compute the $\log L$ from our data on the newly constructed 2D model, as well as the expected
distribution of $\log L$, from simulated stars drawn from the model, as described above.
We stress that the expected distribution of $\log L$ is not unique, but varies for each assumed value of $\sigma_{\rm age}$, being derived from the 2D model computed for the exact value of $\sigma_{\rm age}$

For every value of $\sigma_{\rm age}$ we compare the ``observed'' $\log L$ (i.e. derived from the data) with the predicted one. As mentioned earlier, if the first is significantly smaller than the second, it indicates a poor fit; viceversa, a significantly larger value implies a suspiciously good fit, analogous to a $\chi^2\ll1$. The standard deviation of the predicted $\log L$ distribution allows us to quantify these differences in a meaningful way. This is simplified by the fact the predicted $\log L$ is normally distributed, due to the central limit theorem, since $\log L$ is the sum over a large sample of stars of the individual single-star likelihoods $\log L_i$.

In Figure \ref{fig:improvement_in_logl_with_spread} we present the difference between the observed and predicted $\log L$, normalized on the width of the predicted distribution, as a function of the assumed age spread $\sigma_{\rm age}$, for the two families of evolutionary models and assuming $f=0.5$. The values of $\sigma_{\rm age}$ for which the difference in the y-axis is zero provide the best-fit value of the age spread, the and the projection on the x-axis of the $\pm 1\sigma$ (dotted lines in the Figure), the associated uncertainty.
We find that, for {\sc FRANEC} models the measured age spread is $\sigma_{\rm age}=1.8$~Myr (FWHM$\simeq 4.2$~Myr), while for Siess models this decreases to $\sigma_{\rm age}=1.2$~Myr (FWHM$\simeq 2.8$~Myr). In both cases the uncertainty in $\sigma_{\rm age}$ is about 0.2~Myr.

The observed CMD, with the best-fit 2D density models from {\sc FRANEC} ({\em left}) and {\sl Siess} ({\em right}) grids of isochrones
superimposed is shown again for $f=0.5$ in Fig.~\ref{fig:2Disochrone_example_bestspread}, but with the best-fit 2D models constructed
by including the measured age-spread of 0.5~dex. In this figure is evident how the model density probabilities now follow quite nicely the
observed broadening of the PMS population of LH~95, within the magnitude range $22\leq m_{\rm 555} \leq 27$, which is considered for
the evaluation of the age and age-spread of the system. The difference of these plots with those of Figure \ref{fig:2Disochrone_example_best}, where the best-fit 2D models were constructed assuming coeval PMS populations, is significant. By adding an age spread, the model distributions in the CMD now reproduce well the observed sequence.

Our method predicts a 50\% higher age spread assuming {\sc FRANEC} models than assuming Siess models, and the difference is larger than the uncertainties of each of two values. This is a consequence of the particular shape of the {\sc FRANEC} young isochrones for the lowest masses, where the modeling of the deuterium burning leads to a brighter predicted luminosity. This can be seen already in the original isochrones in the H-R diagram (Figure \ref{fig:plot_theor_isoch}), and is responsible for the over-density seen in both Figure \ref{fig:2Disochrone_example_best}a and \ref{fig:2Disochrone_example_bestspread}a at $V-I\sim 3$ and $I\sim 24$. In this region of the CMD the models predict more stars than actually observed, regardless the assumed age spread. Since the observed population is, therefore, statistically populates more the positions of the CMD where the density of the 2D models is lower, the measured likelihood turns out to be smaller than the predicted one even for the ``right'' age spread. This leads to an overestimation of the value of $\sigma_{\rm age}$.

We remark that this is a general behavior our the method we have proposed: our
technique may overestimate the overall age-spread. This occurs in the case where the isochrones do not follow the observed
sequence, but have a systematically different shape with respect to it.
In fact, the condition we imposed to constrain the measured $\sigma_{\rm age}$ is based on the requirement that our data and a simulated population of test stars obtained from the best 2D density model (including an age spread) produce similar values of ``global'' likelihood $\log L$. If the 2D models have a shape which strongly differs from the observed sequence, the model would never reproduce the data, but there will be always an assumed value of age spread which leads to similar values of likelihood as for the observed sample. In this case, the derived age spread might be higher than the real one. As we discuss in the next section, we can quantify any systematic differences in the shape of the isochrones with respect to the data, and estimate the correct age spread, by analyze the age properties of the population is smaller luminosity ranges.

\begin{deluxetable}{crrrr}
\tablecaption{Age and age spread for different binary fractions}
\tablehead{\colhead{} & \multicolumn{2}{c}{{\sc FRANEC} Models}  & \multicolumn{2}{c}{{\sl Siess}  Models} \\
\colhead{$f$} & \colhead{age} & \colhead{$\sigma_{\rm age}$} & \colhead{age} & \colhead{$\sigma_{\rm age}$} \\
\colhead{} & \colhead{(Myr)} & \colhead{(Myr)} & \colhead{(Myr)} & \colhead{(Myr)} }
\startdata
$f=0.4$ & $3.6$ & $1.4\pm0.2$ & $3.6$ & $1.0\pm0.2$ \\
$f=0.5$ & $3.8$ & $1.8\pm0.2$ & $3.9$ & $1.2\pm0.2$ \\
$f=0.8$ & $4.1$ & $2.3\pm0.25$ & $4.0$ & $1.4\pm0.2$ \\
\enddata
\label{table:agespread_function_of_f}
\end{deluxetable}

We investigated how the assumed binary frequency $f$ affects the derived age spread. We have assumed $f=0.4$ and $f=0.6$ and reapplied the method described above. We stress that, although we previously described our results for a larger range of assumed binary ratios, in the mass range relevant for our study, the value of $f$ is not expected to differ much from 50\% \citep{lada06}.
The results are presented in Table \ref{table:agespread_function_of_f}.
We find that a higher fraction of binaries leads to a larger derived age spread, and, on the contrary, a lower $f$ to a shorter $\sigma_{\rm age}$. Both the results are a consequence of the dependence of the average cluster age as a function of $f$: given that the distance between consecutive isochrones decreases with age. A given spread in the CMD results in a larger age spread if the central isochrone is older, and viceversa.

\subsection{Luminosity-dependence of the age and age-spread}
\label{section:v-bin_ages}

\begin{deluxetable}{ccrcr}
\tablecaption{Variation of ages and age-spreads with luminosity for assumed binarity $f=0.5$}
\tablehead{
\colhead{} &
\multicolumn{2}{c}{{\sc FRANEC} Models}  &
\multicolumn{2}{c}{{\sl Siess}  Models} \\
\colhead{magnitude} &
\colhead{age} &
\colhead{$\sigma$ age-spread} &
\colhead{age} &
\colhead{$\sigma$ age-spread} \\
\colhead{range} &
\colhead{(Myr)} &
\colhead{(Myr)} &
\colhead{(Myr)} &
\colhead{(Myr)}
}
\startdata
$22\leq m_{\rm 555}\leq 23$ &  $0.9$ &  $2.2\pm0.7$ &  $1.0$ &  $2.8\pm0.7$ \\
$23\leq m_{\rm 555}\leq 24$ &  $1.4$ &  $1.7\pm0.5$ &  $1.8$ &  $2.4\pm0.9$   \\
$24\leq m_{\rm 555}\leq 25$ &  $2.6$ &  $0.6\pm0.6$ &  $2.7$ &  $0.9\pm0.8$   \\
$25\leq m_{\rm 555}\leq 26$ &  $3.6$ &  $0.8\pm0.4$ &  $3.6$ &  $0.6\pm0.4$   \\
$26\leq m_{\rm 555}\leq 27$ &  $4.6$ &  $1.0\pm0.2$ &  $4.7$ &  $0.0\pm0.8$
\enddata
\label{table:V-bin_ages}
\end{deluxetable}

We want to verify the presence of
systematic changes of both age and age-spread with luminosity. This effect, which can be caused by uncertainties in the theoretical evolutionary models, as well as in our modeling of the broadening effects in the CMD, can lead to an overestimation of the measured age spread.
To demonstrate this dependence we divide
the photometric sample in the considered magnitude range of $22\leq m_{\rm 555} \leq 27$ into five sub-ranges
1~mag wide each, and we repeat the method described in \S~\ref{section:deriving_cluster_age} to derive the best-fit
age within each subsample. We then apply the method of \S~\ref{section:deriving_age_spread} to derive the
corresponding age-spreads, simulating a family of 2D models with different assumed $\sigma_{\rm age}$ around the
average age measured for each magnitude interval.

The results are given in Table~\ref{table:V-bin_ages}, again for both the {\sc FRANEC} and {\sl Siess} families of evolutionary
models for an assumed binary fraction $f=0.5$.
We find a clear variation of both average age and measured age spread with luminosity.
In particular, for the bright members, close to the PMS-MS transition, which
correspond to masses in the range $1\lesssim M/{\rm M}_{\odot}\lesssim 2$, the derived ages are systematically younger
than those for fainter luminosities (and lower masses). The differences in age are of the order of 4~Myr.
Also, the measured age-spread tends to be significantly larger than the average one for bright stars, and smaller for the lowest masses.
In particular, for $22\leq m_{\rm 555}\leq 23$ the age-spread derived with our method is larger than $\sigma=2.5~Myr$, corresponding to a FWHM exceeding 5~Myr. Considering the very young ages measured in this luminosity interval, and the fact that, for our modeling, the minimum age included in our simulation is 0.5~Myr, this finding implies that the best-fit age distribution for the luminous part of the PMS sequence is highly skewed, with a peak at $\sim 1~Myr$ and a broad tail towards older ages.
On the other hand, for the faintest end of our stellar sample, our method finds that a more modest age spread is needed to produce a simulated distribution of stars as broad as the observed sequence.

The differences in the derived age spreads as a function of luminosities may be due to an incorrect modeling of the 2D isochrones we have used. In particular, some of the broadening effects we included in the model might be mass-dependant. For example, variability and accretion might affect more the brightest members, and less the smallest masses, whereas we assumed identical distributions of these effects along the entire mass range. Considering our results, and averaging the age spreads at different luminosities, the average age spread for the entire stellar sample, derived in Section \ref{section:deriving_age_spread2} should be a fairly well approximation of the real spread.

If we trust our models, the age-luminosity dependance we measure implies that stars in the low-mass regime have been formed some Myrs
{\em before} the intermediate-mass members. However, this is not what is observed in young Galactic 1-2~Myr-old associations,
where no evidence of highly biased IMF are found \citep{hartmann2003}. An explanation for this age difference
among different stellar masses is the assumed binarity. An overestimated binary fraction can lead to overestimation of the age for
the low-mass stars, because the large numbers of unresolved low-mass binaries produce systematically wider modeled distributions
of low-mass stars in the synthetic CMDs. We have investigated this effect by computing the average age in the five selected magnitude
intervals (as those in Table \ref{table:V-bin_ages}), but with 2D modeled distributions constructed assuming no binaries ($f=0$). While
in this case the mass-age dependance reduces, {\sl it does not vanish}, with a reduced variation in the derived ages of the order of
1~Myr. Naturally, this is an extreme case since such a very low binary fraction is not realistic, but unfortunately the true binary fraction
in the considered mass ranges cannot be directly resolved at the distance of the LMC\footnote{At the distance of the LMC, 1 ACS/WFC
pixel corresponds to $\sim2400$~AU.}.

We have investigated more in detail if the mass-age correlation we have found can be a biased result to an inaccuracy in the modeling of the 2D isochrones. In particular, since the effect caused by accretion veiling is the most uncertain among those we have considered (mostly because of the lack of mass accretion estimates for LH~95), we have tried to redo the entire analysis presented in this paper excluding accretion from our modeling. In case of no accretion, we find that the age of the cluster is older (of the order of $6$~Myr, depending on the other parameters), the age spread is larger (several Myrs), and still, there is consistent mass-age correlation, with the very-low mass stars systematically older than the intermediate-mass stars.

Taking into account the above, and if we assume that indeed such a dependence of age with mass should not be observed, as in Galactic
young clusters, we can attribute the observed dependence of age with mass to an inaccuracy of the PMS models we
have utilized. Actually, the inconsistency between the ages predicted by the {\sc FRANEC} and {\sl Siess} isochrones
suggests that the PMS stellar modeling is still far from completely reliable. For a metallicity close to solar,
typical of the Galaxy, several studies have been presented to constrain the evolutionary models based on empirical isochrones
\citep[e.g.,][]{mayne2007}. Certainly new data that sample different young associations and clusters in the metal-poor environment
of the MCs will be exceptionally useful to test the agreement of data with the PMS models also for lower metallicities.

\section{Discussion}
\label{section:discussion}

We stress that our results on the age determination of LH~95 and  the evidence for an age-spread in the system
are valid under the assumption of the correctness of the sources of spread in the CMD considered in our models
(\S~\ref{section:2Disochrones}). It should be noted that having included, apart from photometric uncertainties,
several important biasing factors, namely {\sl differential reddening}, {\sl stellar variability}, {\sl accretion} {\sl binarity} and
{\sl crowding}, our modeling technique is more accurate than similar previous studies, which completely
neglect these biases. Using our models, we cannot reproduce the observed sequence assuming a coeval population, but an additional spread in age is measured.
Nonetheless, as we have shown, our results are very sensitive on the assumptions used to model observed populations in the CMD.
In particular, changes in the assumed binary fraction alter significantly both the average cluster age and the derived age spread.
Also, as discussed previously, the overall effect of accretion veiling influences both this quantities.
For our modeling, we assumed accretion and variability measurements from the Orion Nebula Cluster. If on-going accretion in LH~95 is weaker than we assumed, for instance because of an earlier disruption of the circumstellar disks caused by the flux of the early-type stars present in the region (3 of which are O-type stars, \citep{paperIII}) or simply because of the older age, the measured age would turn out be older and the age spread larger that the results we have presented.

Slow star formation scenarios have been proposed based on both observational and theoretical evidence \citep{pallastahler2000,palla2005,tan2006}; the loose, not centrally concentrated geometry of LH~95, and its relatively low stellar density and mass (see Paper I) would be compatible with a slow conversion of gas into stars \citep{kennicutt1998}. Unfortunately, the absence of dynamical information on our region, as well as a precise estimate of the gas mass do not allow us to infer much about the past evolution of LH~95.

On the other hand, additional unconsidered effects that can displace
PMS stars in the CMD may still be present and affect the derived age-spread; we briefly discuss them here.
The presence of circumstellar material can cause star light to be scattered in the line of sight direction \citep[e.g.][]{kraus-hillenbrand09,guarcello+2010}. This effect may either increase the observed optical flux with respect to that of the central star, or reduce it, if the central star is partially obscured by the circumstellar material. This scenario, however, requires both the presence of a significant circumstellar environment and a particular geometrical orientation, so, if present, would affect only a fraction of the stars in a cluster. In any case, as shown in \citet{guarcello+2010}, stars observed in scattered light appear significantly bluer than the PMS sequence. If such sources are present in our photometric catalog, they would affect only the old tail of our age distribution, or even be excluded as candidate MS field stars. Thus, we are confident that neglecting this effect should not affect significantly our findings.
Another unconsidered effect that could bias the measured age spread is that presented by \citet{baraffe2009}.  Under certain conditions, if a star accretes a significant amount of material it cannot adjust its structure quickly enough, so appears underluminous for a star of its mass. This effect will have a memory, which may persist for several Myr according to the authors, and will cause some young stars to appear older than the are. This would also cause an additional apparent spread in the H-R diagram even in case of a coeval population. The scenario proposed by \citet{baraffe2009}, however, requires episodes of vigorous mass accretion, with a high $\dot{M}$. With no direct measurements of accretion in LH~95, and no information about the past history of accretion, as a function of mass, we are unable to estimate if this scenario could play any role in our region, and if so, to what extent.

\section{Summary}
\label{section:summary}

In this study we utilize the photometric catalog of sources in the young LMC stellar association LH~95, based on HST/ACS
photometry in F555W and F814W filters (corresponding to standard $V$- and $I$-bands). The catalog, which covers stellar
masses as low as 0.2~M$_\odot$ was used in Paper~I to derive for the first time the sub-solar IMF in the LMC. Here we
focus on the age of the system and the determination of the age-spread from the CMD-broadening of its PMS stars.
We first refine the decontamination of our sample from field stars by eliminating any remaining field MS stars
still present in the first catalog of true stellar members (Paper~I). Our photometric catalog of member-stars of LH~95
covers $\sim$~2~000 members in the pre--main-sequence (PMS) and upper--main-sequence (UMS) regions of the CMD.

We develop a maximum-likelihood method, similar to the that described in
\citep{naylor-jeffries06} to derive the age of the system accounting simultaneously
for photometric errors, unresolved binarity, differential extinction, accretion, stellar variability and crowding. The application of the
method is performed in several steps: First, a set of PMS isochrones is converted to 2D probability distributions in the
magnitude-magnitude plane; one model distribution
is constructed for every considered age. A well populated
($N>10^6$ stars) synthetic population is thus created by applying the aforementioned sources of displacement in
the CMD -- except for photometric errors -- to the isochrone for each chosen age. A maximum-likelihood method is
then applied to derive the probability, for each observed star to have a certain age, taking its photometric uncertainty
into account. The multiplication of the likelihood functions of all the stars provides an age probability function for the
entire cluster, from where the most probable age of the cluster is derived.

We apply our method to the stellar catalog of LH~95, assuming the IMF and reddening distribution for the system, as
they were derived in Paper~I. We assume distributions of the variations in both $m_{\rm 555}$ and $m_{\rm 814}$ magnitudes
caused by accretion and variability due to dark spots by utilizing data for the Orion Nebula Cluster from the literature. We model age-dependant
probability distributions using two sets of evolutionary models, that by \citet{siess2000}, for the metallicity of the LMC,
and a new grid computed with the {\sc FRANEC} stellar evolution code \citep{chieffi89, deglinnocenti08} for the same metallicity.
We evaluate the accuracy of our age estimation and we find that the method is fairly accurate in the PMS regime, while the
precision of the measurement of the age is lower at higher luminosities and old ages.
Our method cannot constrain the age of UMS stars due to the small dependence of CMD position with age.
Our treatment showed that the age determination is very sensitive to the assumed evolutionary model and binary fraction; the age
of LH~95 is found to vary from $\sim$2.8~Myr to $\sim$4.4~Myr, depending on these factors.
Assuming  $f=0.5$, the average binary fraction in our mass range \citep{lada06}, the best-fit age of the system derived for this binary fraction is determined to be of the order of 4~Myr.
Despite the sparse and clumpy spatial distribution of PMS stars in LH~95, we find no evidence of systematic variations of age with location.

With our method we are also able to assess the true appearance of an age-spread among the PMS stars of LH~95,
as an indication that star formation in the system occurred not spontaneously but continuously. Our treatment allowed
us to disentangle any {\sl real} age-spread from the apparent CMD-spread caused by the physical and observational
biases which dislocate the PMS stars from their theoretical positions in the CMD (see \S~\ref{section:2Disochrones}).
We find that LH~95 hosts an overall age-spread well modeled by a Gaussian distribution with $FWHM=2.8 - 4$~Myr.

We measure separately ages and age spread as a function of luminosity. For the intermediate masses ($M\sim 1$M$_\odot$) we find younger average ages and larger age spread, whereas for low-mass stars ($M\sim 1$M$_\odot$), older ages and more modest age spreads.
We assign this dependence of age with mass not to a real effect but to imperfections of the PMS evolutionary models, which tend to
predict lower ages for the intermediate masses, and higher ages for low-mass stars. In any case, under the validity of the assumptions used in our modeling, we rule out the possibility that the population of LH~95 is coeval.

Finally, we remark that, despite the method we have presented provides an exceptional tool for an unbiased investigation of ages and age spreads in young PMS regions, the results are highly dependent on the modeling of the assumed sources of apparent spread in the CMD. Only a precise characterization of each of these effects, in different regions, will enable us to give the final answer on the long-standing questions about the duration of star formation.
\\

\acknowledgements
N.D.R. and D.A.G. kindly acknowledge financial support from the German Aerospace Center (DLR) through grants 50OR0401 and 50OR0908 respectively. \\


\end{document}